\documentclass[12pt]{article}
\usepackage{algorithm}
\usepackage{algorithmic}
\usepackage{latexsym}
\usepackage{booktabs}
\usepackage{multirow}
\usepackage{cite}
\usepackage{enumitem}
\usepackage{amssymb,mathrsfs}
\usepackage{booktabs}
\usepackage[justification=centering]{caption}
\usepackage{subfigure}
\usepackage{float}
\usepackage{color}
\usepackage{amsmath}
\usepackage[colorlinks,
            linkcolor=blue,
            anchorcolor=red,
            citecolor=green]{hyperref}
\usepackage{bm}
\usepackage{graphicx}
\usepackage{float}
\usepackage{subfigure}
\usepackage{wrapfig}
\usepackage{amsthm}
\usepackage{epstopdf}
\usepackage{multirow}
\usepackage[table,xcdraw]{xcolor}
\usepackage{booktabs}

\allowdisplaybreaks[4]

\renewcommand{\baselinestretch}{1.0}

\begin{document}
\title{HCGR: Hyperbolic Contrastive Graph Representation Learning  for Session-based Recommendation}

\author{Naicheng Guo\thanks{Ant Group, Beijing, China. Email:\{guonaicheng.gnc, liuxiaolei.lxl, lishaoshuai.lss\}@alibaba-inc.com, \{qiongxu.mqx, yunan.zya, hanbing.hanbing, jefflittleguo.gxb\}@antgroup.com. }
\and
Xiaolei Liu$^\ast$
\and
Shaoshuai Li$^\ast$
\and
Qiongxu Ma$^\ast$
\and
Yunan Zhao$^\ast$
\and
Bing Han$^\ast$
\and
Lin Zheng\thanks{Department of Computer Science, Shantou University, Shantou, China. Email:lzheng@stu.edu.cn. }
\and
Kaixin Gao \thanks{School of Mathematics, Tianjin University, Tianjin, China.
Email: gaokaixin@tju.edu.cn.}
\and
Xiaobo Guo$^\ast$
}

\date{}

\maketitle

\begin{abstract}
\noindent

\vspace{3mm}
  Session-based recommendation (SBR) learns users' preferences by capturing the short-term and sequential patterns from the evolution of user behaviors. Among the studies in the SBR field, graph-based approaches are a relatively powerful kind of way, which generally extract item information by message aggregation under Euclidean space. However, such methods can't effectively extract the hierarchical information contained among consecutive items in a session, which is critical to represent users' preferences. In this paper, we present a hyperbolic contrastive graph recommender (HCGR), a principled session-based recommendation framework involving Lorentz hyperbolic space to adequately capture the coherence and hierarchical representations of the items. Within this framework, we design a novel adaptive hyperbolic attention computation to aggregate the graph message of each user's preference in a session-based behavior sequence. In addition, contrastive learning is leveraged to optimize the item representation by considering the geodesic distance between positive and negative samples in hyperbolic space. Extensive experiments on four real-world datasets demonstrate that HCGR consistently outperforms state-of-the-art baselines by 0.43$\%$-28.84$\%$ in terms of $HitRate$, $NDCG$ and $MRR$.

\vspace{3mm}

\noindent {\bf }\hspace{2mm}

\end{abstract}

\section{Introduction}
In many e-commerce online scenarios, user profiles usually cannot be obtained so that session-based recommendation has become an important solution to address anonymous recommendation. Session-based recommender (SBR) system  learns users' preferences by mining sequential patterns from users' chronological historical behavior without user profiles to predict users' future interests in one session.
Most traditional Markov chain based SBR models, e.g., FPMC \cite{rendle2010factorizing}, FOSSIL \cite{he2016fusing}, conduct sequence modeling and prediction only by considering user's last behavior.
Lately, RNN-based models the treat historical behaviours of each user as a strictly-order, temporally dependent sequence like linguistic sentences. These methods like GRU4REC \cite{hidasi2015session}  surprisingly increase the performance in many real scenarios because of their effectiveness in storing short-term information. However, they assume that the adjacent items in a session have a fixed sequential dependence which is unable to capture the user interest variations, as a result,  they are prone to introduce wrong dependencies.
To address this issue, later models that fusing self-attention mechanisms like SASREC, were proposed. Guo et al. \cite{zheng2020sentiment}
 further improve the attention-based approaches by introducing specialized human sentiment factors.
The aforementioned attention-based methods prefer to model unidirectional message transformation between adjacent items in a sequence. Such transformation may lose the insight of the relevant information of the whole sequence.
For example, in a music player APP, although a user may randomly play an album or a certain type of music, which will generate different playback records, it does not mean that the user's interest has changed. In other words, strictly modeling the user's local click records and ignoring the global relationship may lead to overfitting.
To resolve the problem in attention-based methods, GNN-based models like SR-GNN \cite{wu2019session} and GC-SAN \cite{xu2019graph} utilize graphs to capture the coherence of items within a session due to their powerful ability to represent structured data and adopt attention layers to learn long-term dependence. Despite the leading performance of GNN-based models compared with traditional SBR methods, great challenges remain.\\

\textbf{Challenge1:}
The user's interests are extensive and hierarchical, which can be expressed as a power-law distribution of items clicked by users. The existing session-based recommendation methods learn the representations in Euclidean space, which can't effectively capture the information of such hierarchical, or in other words, tree-structured data.

\textbf{Challenge2:}
Recent studies have proved that hierarchical data can be better explained under Non-Euclidean geometry of low-dimensional manifolds. But in the GNN-based methods, introducing Non-Euclidean transformation will result in the discrepancy between Euclidean and Non-Euclidean space when aggregating neighbouring messages and applying attention mechanism.
\\

\begin{figure}[htbp]
\centering
\begin{minipage}[t]{0.4\textwidth}
\centering
\includegraphics[width=4cm]{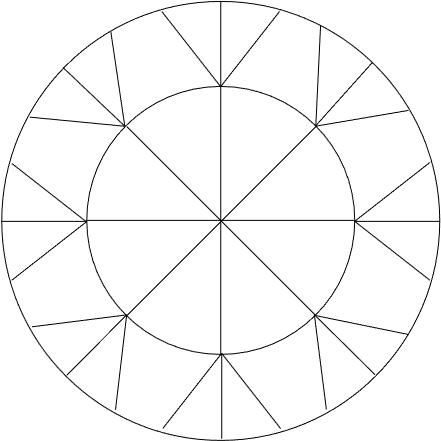}
\caption{Euclid space}
\label{Euclid}
\end{minipage}
\begin{minipage}[t]{0.4\textwidth}
\centering
\includegraphics[width=4cm]{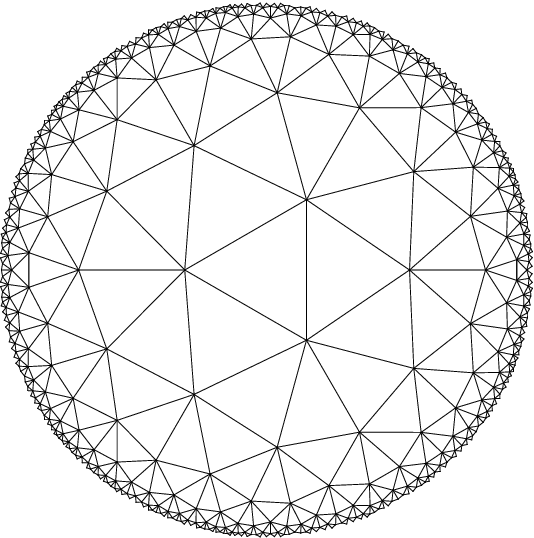}
\caption{2D Poincare disk}
\label{Poincare}
\end{minipage}
\end{figure}

These challenges remain prevalent in real-world recommendation scenarios since it has been demonstrated that users' behaviors like clicking or purchasing have prototypical characteristics of complex structures, which are generally power-law distributed \cite{ravasz2003hierarchical, krioukov2010hyperbolic, papadopoulos2012popularity}. Aforementioned, data with hierarchical structure could be well represented in hyperbolic space. This has motivated representation learning in hyperbolic space to effectively capture the information of the user behaviors with hierarchical property \cite{sun2021hgcf}. Furthermore, the hyperbolic representations can naturally capture the similarity and hierarchy by their distance. To illustrate the difference between Euclidean and hyperbolic space, we visualized them in Figure \ref{Euclid} and Figure \ref{Poincare}. In a two-dimensional Euclidean space as showed in Figure \ref{Euclid}, the number of nodes increases polynomial in the center to the radius. By contrast, in two-dimensional hyperbolic space in Figure \ref{Poincare}, the number of nodes increases exponentially in the center to the given radius, and the hyperbolic space has more powerful (exponential-level) representation ability than then Euclidean space\cite{Asaduzzaman2020HolographyOT}. In conclusion, given the same radius, hyperbolic space has larger space, thereby including more nodes. Therefore, the general representational capacity of Euclidean space can be summarized as a square level, which can cause high distortion if we model the data of hierarchical relational users' preferences.

To overcome these session-based challenges above, in this paper we propose a novel graph neural network framework, namely Hyperbolic Contrastive Graph Recommender (HCGR), upon hyperbolic space, specifically, Lorentz hyperbolic space for its simplicity and stability, to optimize the underlying hierarchical embeddings. First, we embed items with dense and effective representations, which are predisposed to preserving their internal hierarchical properties in Lorentz hyperbolic space. To ensure the correctness of the necessary representations' transformation, we utilize some specific operations based on the  Lorentz hyperboloid model. Second, we construct an improved graph neural network framework and a novel message propagation mechanism to model the preferences in user behavior sequences. To enable the model better distinguish users' preferences for different items, we propose an adaptive hyperbolic attention calculation method. Third, we introduce contrastive learning to optimize the model by considering the distance between positive and negative samples in hyperbolic space.

Overall, we summarize our main contributions of this work as follows:
\begin{itemize}
\item  We exploit hyperbolic item representation for session-based recommendation, to the best of our knowledge, our method is the first one to extract hierarchical information of user behaviors within a session under hyperbolic space for SBR tasks.
\item  We design a novel attention calculation approach in hyperbolic space to deal with graph information aggregating, which can't be effectively implemented by the existing aggregation methods in Euclidean space.
\item  We introduce contrastive learning to optimize the item representation by considering the geodesic distance between positive and negative samples in Lorentz hyperbolic space.
\item We conduct extensive experiments on three public datasets and one financial service industrial dataset, which show that our Lorentz hyperbolic session-based recommendation framework can achieve better performance compared to the state-of-the-art SBR methods in terms of $HitRate@K$, $NDCG@K$ and $MRR@K$, where $K=${$10,20$}.
\end{itemize}

The structure of this paper is as follows. Section \ref{related_work} presents the related work, including session-based recommendation and hyperbolic learning. Section \ref{preliminaries} introduces the preliminary work of this paper, including the definition of the problem and some basic knowledge of the Lorentz hyperbolic space. Section \ref{methodolodgy} introduces the implementation details of HCGR framework, and section \ref{experiment} gives the experimental results. Finally, conclusions are set out in section \ref{conclusion}.

\section{Related work}
\label{related_work}
In section \ref{session_based_rec} we review a line of representative works on session-based recommendation, including traditional MCs(markov chains) based models, RNNs(recurrent neural networks) based models, attention-based models, and GNNs(graph neural networks) based models. Then, in section \ref{hyperbolic_learning} we discuss related hyperbolic representation learning methods related to our proposed HCGR.

\subsection{Session-based Recommendation}
\label{session_based_rec}
\subsubsection{Markov chain models}
Early sequential recommendation methods mainly rely on Markov chains. For example, FPMC \cite{rendle2010factorizing} combines MF(matrix factorization) and MC to the learn general preference and local interest of users for the next basket recommendation. HRM \cite{wang2015learning} applies non-linear operations to extract more complex pattern of both user's sequential behavior and interests. Fossil \cite{he2016fusing} fuses similarity-based methods with Markov chains to conduct personalized sequential recommendations. A shortcoming of MC-based models is that it is difficult to learn long-term dependencies because MC-based models assume that the next state is only related to the previously nearest state. Although some high-order Markov models can associate the next state with several previous states, they consume high computational cost\cite{liu2018stamp,wu2019neural}.

\subsubsection{Recurrent neural networks models}
In recent years, researchers adopt RNNs to capture time dependency in temporal data. The first RNN-based sequential recommendation method is GRU4REC \cite{hidasi2015session}, which uses GRU(gated recurrent unit) to capture long-term dependencies among sessions. Leveraging a novel pair-wise ranking loss, GRU4Rec \cite{hidasi2015session} significantly outperforms MC-based approaches. Inspired by GRU4REC \cite{hidasi2015session}, MV-RNN \cite{cui2018mv} incorporates visual and textual information to alleviate the item cold start problem. Furthermore, ROM \cite{zheng2020memory} utilizes an interactive self-attention mechanism to adaptively reorganize the entity memory and the topic memory for the rating prediction task. However, RNN-based methods assume that the adjacent items in a session have a fixed sequential dependence, which may generate wrong dependencies and introduce noises in real-world scenarios like music playing.

\subsubsection{Attention mechanism}
Recent models with attention mechanism \cite{vaswani2017attention} perform particularly well in sequential recommendation. Li et al. explores a hybrid encoder with attention mechanism to model users' sequential behaviors and user's interests in the same session \cite{li2017neural}. A short-term attention priority model STAMP \cite{liu2018stamp} is proposed, which can capture both personal interest from the long-term memory of the session context, and user's current interest from the short-term behaviors. SASREC \cite{kang2018self}  effectively captures users' long-term preferences from the sparse and dense datasets, and FDSA \cite{zhang2019feature}
puts the features of behaviours and items into two distinct independent blocks of self-attention to model the transition patterns of the items and the transition patterns of the items and achieve remarkable effects.

\subsubsection{Graph neural networks}
Most advanced sequential recommendation models apply the self-attention mechanism to capture user behavior relations in a long sequence. However, it is a challenge to find out both implicit and explicit relation between adjacent behaviors.
GNNs can find out such relations effectively \cite{huang2018improving,wu2019neural} and can capture complex interaction of user behaviors. For example, SR-GNN \cite{wu2019session} instructively constructs a digraph for each sequence and CS-SAN \cite{xu2019graph} further incorporates self-attention mechanism to generate the representation of the constructed digraph. In addition, Wu et al. focuses on users in the session and models their historical sequence by dot-attention mechanism \cite{wu2019personalizing}. FGNN \cite{qiu2019rethinking} proposes a weighted attention layer and a readout function to learn item embeddings and session embeddings for next item recommendation. Recently, Ma et al. utilizes memory models to capture both the long term and short-term user behaviors  \cite{ma2020memory}. Wang et al. introduces a new GNN-based method that can learn global relations between items \cite{wang2020beyond}. Chen et al. \cite{chen2020handling} utilizes several ways to reduce the information within message propagation.
GNN-based methods have yielded many fruitful results, but the existing methods generally model user behavior in tangent space, and the representation learned in tangent space is limited to capturing attributes of shallow properties and lack of hierarchy. In this work, we aim to learn hierarchical item representation in Lorentz hyperbolic space, and we want to find deep user behaviors patterns in session-based recommendation.

\subsection{Hyperbolic Learning}
\label{hyperbolic_learning}
Recently, many studies have shown that complex data may exhibit a highly non-Euclidean structure \cite{Br2017,Du2010}. Researchers are increasingly considering building complex neural networks on Riemannian space, in which the hyperbolic space with negative constant curvature is an attractive option \cite{Ya2017,Br2006}. In many domains, such as sentences in natural language \cite{Al2019}, social networks \cite{Ya2017}, biological protein graph \cite{Br2006}, etc., data usually have a tree-like structure or can be represented hierarchically, and hyperbolic space is equipped to model hierarchical structures naturally and effectively \cite{Ma2017}. Due to its strong representation ability, hyperbolic space has been applied in many application areas \cite{qi2019,In2019,Bo2020,Nu2021,sh2020}. For instance, Liu et al. \cite{qi2019} proposed Hyperbolic Graph Convolutional Networks used for graph representation learning by combining the expressiveness of hyperbolic space and Graph Convolutional Networks. Chen et al. \cite{Bo2020} proposed a hyperbolic interaction model for multi-label classification tasks. These works have shown the advantages and effectiveness of hyperbolic space in learning hierarchical structures of complex relational data.

Noticing the potential of hyperbolic space in learning complex interactions between users and items, many researchers have tried to apply hyperbolic learning to recommendation systems \cite{be2019,Ky2020,ch2021,An2021,ob2018}. Chamberlain et al. \cite{be2019} proposed a large-scale recommender system based on the hyperbolic space, which can be scaled to millions of users. \cite{Ky2020} considered constructing multiple hyperbolic manifolds to map the representation of user and ad, and proposed a framework that can effectively learn the hierarchical structure in users and ads based on the hyperbolic space. Ma et al. \cite{ch2021} proposed a recommendation model in the hyperbolic space for Top-K recommendation. Li et al. \cite{An2021} presented the Hyperbolic Social Recommender which utilized hyperbolic geometry to boost the performance. Wang et al.  \cite{wang2021hypersorec} proposed a novel graph neural network framework (HyperSoRec) combing hyperbolic learning for the social recommendation.

\section{Preliminaries}
\label{preliminaries}
In this section, we introduce basic knowledges about hyperbolic geometry and graph neural networks works related to our proposed HCGR.

\subsection{Hyperbolic Geometry}
Hyperbolic space is a Riemannian surface with negative curvature. Several hyperbolic geometric models have been widely used, including Poincare disk model \cite{ob2018,Al2019}, Klein model \cite{c2019} and Lorentz (hyperboloid) model \cite{ma2018}. All these hyperbolic models are isometrically equivalent, i.e., any point in one of these models can be transformed to another point with distance-preserving transformations \cite{1995Introduction}. In this paper, we choose the Lorentz model as the framework cornerstone, because of the numerical stability and calculation simplicity of its exponential/logarithm maps and its distance function.
In hyperbolic geometry, we use the Lorentz formulation to model the network, which is found to be more stable for numeric optimization patterns \cite{ma2018}. We want to learn $d$-dimensional user and item embeddings.

A $d$-dimensional hyperbolic space is a Riemannian manifold $\mathcal{M}$ with a constant negative curvature, which is denoted by $c$. The negative reciprocal of the curvature is denoted by
$
k=-\frac{1}{c},
$
where $k>0$. The Lorentz representation is defined by the pair $\mathcal{L}^d=(\mathcal{H}^d, g_{\mathcal{L}})$ and
\begin{equation}
    \mathcal{H}^d=\left\{
    \mathbf{x} \in \mathbb{R}^{d+1}|\langle \mathbf{x},\mathbf{x}\rangle_{\mathcal{L}}=-k, \mathbf{x}_0>0
    \right\},
\end{equation}
where $\langle \mathbf{x},\mathbf{y}\rangle_{\mathcal{L}}$ is the Lorentz inner product given by
\begin{equation}
\langle \mathbf{x},\mathbf{y}\rangle_{\mathcal{L}}=-\mathbf{x}_0\mathbf{y}_0+
\sum\limits_{i=1}^d\mathbf{x}_i\mathbf{y}_i, \quad
\forall~ \mathbf{x},\mathbf{y}\in\mathbb{R}^{d+1},
\end{equation}
and the metric matrix $g_{\mathcal{L}}$ is given by
\begin{equation*}
g_{\mathcal{L}}=\left[
\begin{array}{cccc}
  -1 &  &  &  \\
   & 1 &  &  \\
   &  & \ddots &  \\
   &  &  & 1
\end{array}
\right].
\end{equation*}
The distance function induced by the metric $g_{\mathcal{L}}$ is
\begin{equation}
d_\mathcal{L}(\mathbf{x},\mathbf{y})=\sqrt{k}\text{arcosh}
\left(
-\frac{\langle \mathbf{x},\mathbf{y}\rangle_\mathcal{L}}{k}
\right).
\end{equation}
For any pair of points $\mathbf{x},\mathbf{y}\in\mathbb{R}^{d+1}$, the tangent space $\mathcal{T}_\mathbf{x}\mathcal{H}^d$ at point $\mathbf{x}$ is a $d$-dimensional Euclidean space. The elements of $\mathcal{T}_\mathbf{x}\mathcal{H}^d$ are referred to as tangent vectors and satisfying
\begin{equation}
\mathcal{T}_\mathbf{x}\mathcal{H}^d=\{\mathbf{v}\in\mathbb{R}^{d+1}| \langle \mathbf{v},\mathbf{x} \rangle_\mathcal{L}=0\}.
\end{equation}

\subsection{Graph Neural Network}
GNNs are neural networks that can handle graph-structured data directly. They are often applied in  classification, link prediction and graph classification tasks. In this paper, we focus on graph classification, because we formulize each user's behavior to a graph and we want to learn a representation from it rather than a single node.

Let $G(V,E)$ denotes a given graph, where $V$ and $E$ are the set of the nodes and edges respectively, where $x_v$ represents the feature vector of $v\in V$, which is the initial embedding of node $v$. To specific, we formulate the graph classification task as follow. Our work is to learn a classifier $f$ and the graph-level representation $H$ to predict the label of the graph. Given a collection of graphs ${(G_1,G_2,\ldots,G_n)}\in G$ and the corresponding labels ${(v_{L_1},v_{L_2},\ldots,v_{L_n})} \in V_L$.

GNNs use the structure of graph and the original feature of each node to learn its corresponding representation. The learning process is to take a node as the center, and iteratively aggregate the neighborhood information along edges. The information aggregation and update process can be formulated as follows:

\begin{equation}
       {\mathbf{t}_v^{(l+1)}}=f_{aggregator}(\mathbf{x}_u^{(l)},u\in N(v)),
\end{equation}
\begin{equation}
        \mathbf{x}_v^{(l+1)}=f_{updater}(\mathbf{x}_v^{(l)}, {\mathbf{t}_v^{(l+1)}}),
\end{equation}
where ${\mathbf{x}_v^l}$ represents the embedding of node $v$ after $l$-th layer aggregator and $N(v)$ is neighborhood of node $v$. The information aggregation function $f_{aggregator}$ aggregates the information from the neighborhood information and passes it to the target $v$. The update function $f_{updater}$ calculates the new node statues from the source embedding ${{\mathbf{x}_v^l}}$ and the aggregated information ${\mathbf{t}_v^{l+1}}$.

After $l$ steps of information aggregation, the final embedding gather the $l$-hop neighborhood and the structure information. For the graph classification task, readout function $f_{readout}$ generates a graph level embedding $\mathbf{Z}$ by gathering the embeddings of all node in the final layers:
\begin{equation}
        \mathbf{Z} =f_{readout}(\{{\mathbf{x}_v^{(l)},v\in V}\}).
\end{equation}

\section{Methodolodgy}
\label{methodolodgy}
\begin{figure}[ht]
\centering
\includegraphics[width=1.0\textwidth]{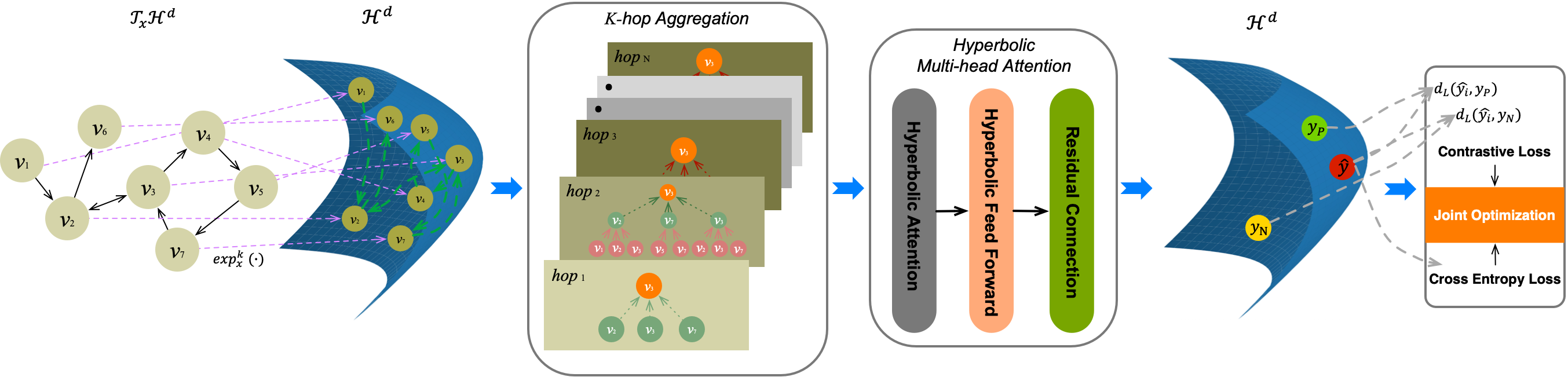}
\caption{The Architecture of the HCGR Framework. It takes user' historical sequential items as input, and then build the directed graph being projected to the hyperbolic space, finally outputs the predicted probability of the next item that the user most likely to click.}
\label{Fig:NetworkArch}
\end{figure}

In this section, we describe the implementation details of the HCGR framework. First, in section \ref{notation_and_problem_def} we illustrate the notations used in this paper and define the session-based recommendation task. In section \ref{sec:4.2}, we transform the user behaviors within a session into a session graph, and present the HCGR's overall pipeline in Figure \ref{Fig:NetworkArch}. In section \ref{emb_in_hyp}, we introduce the embeddings in Lorentz hyperbolic space. Next, we describe the novel attention mechanism that is especially designed for hyperbolic geometry in section \ref{hyp_graph_att}. After learning the embeddings, we set up the hyperbolic attention mechanism to construct representation of user behaviors (section \ref{hyp_att}).Finally, we describe the contrastive learning with hyperbolic space distance (section \ref{contrastive_loss}).

\subsection{Notation and Problem Definition}
\label{notation_and_problem_def}
A session-based recommendation task is constructed on historical user behavior sessions, and makes predictions based on current user sessions. In this task, there is an item set $V$, where $m=|V|$ is the number of items and all items are unique. Each session $S=[v_{1},v_{2},\ldots,v_{_n}]$ is composed of a series of user's interactions, where $v_{i}$ represents an item clicked at the $i$-th position in $S$ and $n$ represents session's length. Our purpose of session-based recommendation is to predict the item that the user is most likely to click on next time in a given session $S$.

For each given session $S$ in the training process, there is a label as the target. In the training process, for each item $v_{{i}}\in V$ in given session, our model learns the corresponding embedding vector $\mathbf{x}_v\in \mathbb{R}^d$, where $d$ is the dimension of vector $\mathbf{x}_v$. Our model outputs a probability distribution $\hat{\mathbf{y}}$ over the given session $S$, where the item with Top-$K$ value will be regarded as the candidate for Top-$K$ recommendation.

\begin{table}[]
\centering
\caption{The key mathematical notations}
\begin{tabular}{cc}
\toprule
Notation & Description \\
\midrule
 $\mathcal{H}^d$  & a hyperbolic space of dimension $d$ \\
 $\mathcal{M}$               & Riemannian manifold  \\
 $S$               &    a given session        \\
 $c$               & the curvature of hyperbolic space \\
 $k$               & the negative reciprocal of the curvature $c$  \\
 $\mathcal{T}_\mathbf{x}\mathcal{H}^d$               & the tangent space at point $\mathbf{x}$ with dimension $d$ \\
 $x_v$             &   a item embedding in the Euclid space  \\
 $l$               &    the layer of graph neural network    \\
 $\mathbf{x}^H$               &    a item embedding in the hyperbolic space  \\
\bottomrule
\end{tabular}
\end{table}

\subsection{Behaviors Graph}
\label{sec:4.2}
\begin{figure}[ht]
\centering
\includegraphics[width=0.6\textwidth]{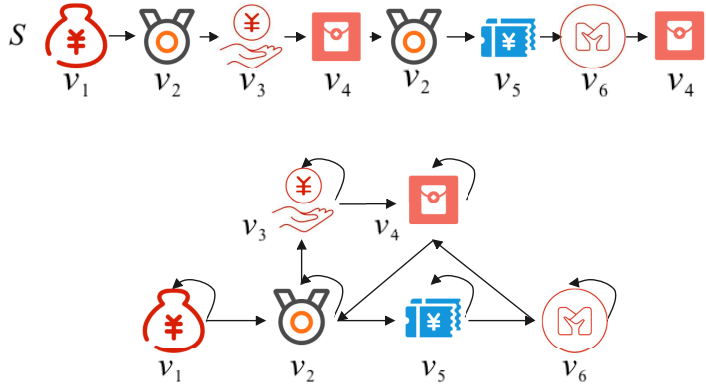}
\caption{An example of how to convert a session to graph}
\label{Fig:aaa}
\end{figure}

Because graph neural network can't deal with session directly, the first thing we need to do is converting a given session $S=[v_1,v_2,\ldots,v_n]$ to the session graph $G_s$. According to the analysis of datasets, it is very common for users to click the same item multiple times within the session. Because the user's behavior is chronological and the same item may be clicked multiple times, we choose the weighted directed graph to represent the changing process of the given session $S$. All the sessions will be converted into session graphs. To show this process more clearly, we show the process of this session converter in Figure \ref{Fig:aaa}. We use $E_s$ to denote all weighted directed edges set. Its elements are composed of $(v_t,v_{t+1},w_{t,t+1})$, where $v_t$ , $v_{t+1}$ is the item clicked at timestamp $t$,${t+1}$ respectively, $w_{t, t+1}$ denotes the weighted directed edge between $v_t$ and $v_{t+1}$. Note that if the node does not have a self-loop, we will add a self-loop with weight $1$ to it. Each node represents the unique item in the session and the features $\mathbf{x}_v$ are initialized in the Lorentz hyperbolic space which introduced in section \ref{emb_in_hyp}.

\subsection{Embeddings in Lorentz Hyperbolic Space}
\label{emb_in_hyp}
We use the representation from Lorentz hyperbolic space for item embedding. The $k=-\frac{1}{c}$ is the reciprocal of curve $c$, which treated here as a trainable parameter and initials empirically. Then we fix the origin $\mathbf{o}=(\sqrt{k},0,\ldots, 0)\in \mathcal{H}^d$ and use it as a reference point. The embeddings are initialized by sampling the Gaussian distribution on the Euclidean space $\mathcal{T}_\mathbf{x}\mathcal{H}^d$ of the reference point o.

We denote the mapping between hyperbolic spaces and tangent spaces as exponential map and logarithmic map, respectively. The exponential map is a map from subset of a tangent space $\mathcal{T}_\mathbf{x}\mathcal{H}^d$ to $\mathcal{H}^d$. The logarithmic map is the reverse map that maps $\mathcal{H}^d$ back to the tangent space $\mathcal{T}_\mathbf{x}\mathcal{H}^d$. For any $\mathbf{x},\mathbf{y}\in\mathcal{H}^d$ and $\mathbf{v}\in \mathcal{T}_\mathbf{x}\mathcal{H}^d$ satisfying $\mathbf{x}\neq \mathbf{y}$ and $\mathbf{v}\neq 0$, the exponential map $\exp_\mathbf{x}^k(\cdot)$ and logarithmic map $\log_\mathbf{x}^k(\cdot)$ are defined as follows:
\begin{equation}
\label{eq_exp}
\exp_\mathbf{x}^k(\mathbf{v})=\cosh\left(
\frac{\|\mathbf{v}\|_\mathcal{L}}{\sqrt{k}}
\right)\mathbf{x}+
\sqrt{k}\sinh\left(
\frac{\|\mathbf{v}\|_\mathcal{L}}{\sqrt{k}}
\right)
\frac{\mathbf{v}}{\|\mathbf{v}\|_\mathcal{L}},
\end{equation}

\begin{equation}
\label{eq_log}
\log_\mathbf{x}^k(\mathbf{y})=d_\mathcal{L}(\mathbf{x},\mathbf{y})
\frac{\mathbf{y}+\frac{1}{k}\langle \mathbf{x},\mathbf{y}\rangle_\mathcal{L}\mathbf{x}}
{\left\|
\mathbf{y}+\frac{1}{k}\langle \mathbf{x},\mathbf{y}\rangle_\mathcal{L}\mathbf{x}
\right\|_\mathcal{L}} ,
\end{equation}
where $\|\mathbf{v}\|_\mathcal{L}=\sqrt{\langle \mathbf{v},\mathbf{v}\rangle_\mathcal{L}}$ is the Lorentzian norm.

\subsection{Hyperbolic Graph Attention Network}
\label{hyp_graph_att}
Following the mapping layer, how to model session graph and mine user preferences is the key issue. Users typically click on several items they like, and these items have rich hierarchical structure. As a result, we propose a novel information aggregation with attention mechanism in hyperbolic space to capture the influence of different items on user preferences during the process of information propagation. Hyperbolic space can better represent item\cite{be2019,Ky2020}, but we still face a technical challenge, the traditional hyperbolic model does not define the necessary vector operation process, such as vector addition and multiplication etc. Inspired by previous works \cite{be2019,Ky2020,wang2021hypersorec}, we formulate the multiplication and addition operation in hyperbolic space as follow:
\begin{equation}
W\otimes^k \mathbf{x}^H:=\exp_\mathbf{o}^k(W\log_\mathbf{o}^k(\mathbf{x}^H)),
\end{equation}
\begin{equation}
\mathbf{x}^H\oplus^k \mathbf{b}:=\exp_{\mathbf{x}^H}^k(P_{o\rightarrow \mathbf{x}^H}^k(\mathbf{b})),
\end{equation}
where $P_{o\rightarrow \mathbf{x}^H}^k(b)$ is the Parallel Transport: for two point $\mathbf{x},\mathbf{y}$ on the Lorentz space $\mathcal{H}^d$, the parallel
transport of a tangent vector $\mathbf{v} \in \mathcal{T}_\mathbf{x}\mathcal{H}^d$ to the tangent space $\mathcal{T}_\mathbf{x}\mathcal{H}^d$ is:
\begin{equation}
P_{\mathbf{x}\rightarrow \mathbf{y}}(\mathbf{v})=\mathbf{v}-\frac{\langle\log_\mathbf{x}(\mathbf{y}),\mathbf{v}\rangle_\mathcal{L}},
{d_\mathcal{L}(\mathbf{x},\mathbf{y})^2}(\log_\mathbf{x}(\mathbf{y})+\log_\mathbf{y}(\mathbf{x})).
\end{equation}
Non-linear activation with different curvatures is proposed as follow:
\begin{equation}
\sigma^{\otimes^{k_{l+1},k_l}}(\mathbf{x}^H)=\exp_\mathbf{o}^{k_{l+1}}(\sigma(\log_\mathbf{o}^{k_l}(\mathbf{x}^H))),
\end{equation}
where $-\frac{1}{k_{l+1}}$, $-\frac{1}{k_l}$ is the hyperbolic curvature at layer $l$, $l+1$ respectively. To specific, we project the embedding from the tangent space $\mathcal{T}_\mathbf{x}\mathcal{H}^d$ to $\mathcal{H}^d$ via exponential map $\exp_\mathbf{x}^k(\cdot)$ according to the Eq(\ref{eq_exp}). Then we perform the addition and multiplication operation according to the above equation.

The crucial idea of traditional GNNs is to learn representations in the given graph by iteratively aggregating and capture multi-hop neighborhood structures and features. The process of information aggregation usually consists of two parts: feature transformation and nonlinear activation. Recent studies have shown that, compared to the simple average aggregation method, the gain with features transformation and non-linear activation is rarely, and may even bring negative gain. In addition, these two operations may lead to significant over fitting of highly sparse user behavior\cite{he2020lightgcn}. Based on these studies, we remove the unnecessary feature transformation and non-linear activation to accelerate the training, inference and reduce the complexity of our framework.

To make better use of the representation ability of Lorentz space, we redesigned a way of information aggregation in hyperbolic space. We refer to the ideas of GCN\cite{kipf2016semi} and GAT\cite{velivckovic2017graph} to calculate the attention weight between target node and neighbors respectively. The detailed calculation way is shown as follow:
\begin{equation}
\label{wij}
w_{ij} = Softmax_{j\in \mathcal{N}(i)}(W_a(\log_\mathbf{o}^k(\mathbf{x}_i^{H,l})||\log_\mathbf{o}^k(\mathbf{x}_j^{H,l}))+b_a),
\end{equation}
\begin{equation}
\mathbf{x}_i^{H,l+1} = \exp_{\mathbf{x}_i^{H,l}}^k(\sum_{j \in \mathcal{N}(i)}w_{ij}\log_{\mathbf{x}_i^{H,l}}^k(\mathbf{x}_j^{H,l})),
\end{equation}
where $W_a$ is  $\mathbb{R}^{1\times(2d+2)}$ matrices, and $b_a$ is a constant number. The learning process is to take a node $i$ as the center, and iteratively aggregate the neighbors information along edges. For each node $i$, in hyperbolic mechanism, all attention coefficients of their neighbors can be calculated as Eq(\ref{wij}).
In order to use these attention coefficients, a linear combination for the neighbors is used for updating the embeddings of the nodes.

To take full advantage of higher-order relationships, we need to stack multiple hyperbolic attention layers together.

\begin{equation}
\mathbf{z}_i^{H} = \exp_\mathbf{o}^{k_l}(\alpha_0*\log_\mathbf{o}^{k^0}(\mathbf{x}_i^{H,0})+\alpha_1*\log_\mathbf{o}^{k^1}(\mathbf{x}_i^{H,1}) + \ldots+\alpha_l*\log_\mathbf{o}^{k^l}(\mathbf{x}_i^{H,l})).
\end{equation}

\subsection{Hyperbolic Attention Mechanism}
\label{hyp_att}
After we obtain the graph level representation $\mathbf{Z^H}= [\mathbf{z}_1^H, \mathbf{z}_2^H,\ldots,\mathbf{z}_n^H]$, we want to utilize self-attention mechanism to better capture user's preference.
Self-attention is an important part of attention mechanism, it has yielded many fruitful results such as:\cite{vaswani2017attention},\cite{kang2018self},\cite{zhang2019feature}. The self-attention mechanism can calculate the global dependence between user behavior and capture the item transformation relations of the whole session sequences.
The original self-attention mechanism does not define in hyperbolic space, so we extend the self-attention mechanism to hyperbolic space and we formalize the hyperbolic self-attention mechanism as follows:
\begin{equation}
\mathbf{F}^H = \exp_\mathbf{o}^k(Softmax(\frac{(W^Q\log_\mathbf{o}^k(\mathbf{Z}^H))(W^k\log_\mathbf{o}^k(\mathbf{Z}^H))^T}{\sqrt{d+1}}(W^V\log_\mathbf{o}^k(\mathbf{Z}^H)))),
\end{equation}
where $W_Q$, $W_K$ and $W_V$ are $\mathbb{R}^{(d+1)\times(d+1)}$ matrices. It will receive the query ($Q$), key ($K$), and value ($V$), and calculate the similarity between each element in the session through the scaled dot-product, so as to characterize the user's long-term preference, where d is the dimension of the input vector and $\sqrt{d+1}$ is the scale factor, which is used to prevent the gradient vanishing problem caused by the large value after the dot product.Correspondingly, Element-wise Feed-Forward is also extended to hyperbolic space and it is given by:
\begin{equation}
\mathbf{E}^H=\exp_\mathbf{o}^{k_2}(\log_\mathbf{o}^{k_2}(W_2\otimes^{k_2}(\sigma^{\otimes^{k_{2},k_1}}(W_1\otimes^{k_1} \mathbf{F}^H \oplus^{k_1} b_1))\oplus^{k_2} b_2)+\log_\mathbf{o}^{k_1}(\mathbf{\mathbf{F}}^H)),
\end{equation}
where $W_1$ and $W_2$ are $\mathbb{R}^{(d+1)\times(d+1)}$ matrices, $b_1$ and $b_2$ are $(d+1)$-dimensional bias vectors.
It takes full account of the interaction between the dimensions of various vectors through nonlinear activation function and linear transformation. A skip connection after the feed forward network, which makes the model reduce the loss of information and takes advantage of the low-layer information.
For simplicity, we define the entire hyperbolic self-attention mechanism above as:

\begin{equation}
    \mathbf{E}^H =\textit{Hyp-Self-Att}(\mathbf{Z}^H).
\end{equation}

Recent studies have shown that, different layers of self-attention mechanism may capture different types of features, so it is necessary to increase the number of layers appropriately to enhance the model expression. The multi-layer hyperbolic self-attention mechanism is define as:
\begin{equation}
\mathbf{E}^{H,j}=\textit{Hyp-Self-Att}(\mathbf{E}^{H,j-1}).
\end{equation}

Finally, the hyperbolic self-attention mechanism output is $\mathbf{E}^{H,j}$. After $j$-th adaptive hyperbolic self-attention blocks, we obtain the long-term attentive session representation $\mathbf{E}^j$. The short-term interest describes the current preferences of users. It is based on several items recently visited as the basis for prediction. The next behavior of users is often closely related to his recent interests. In order to better model the relationship in the whole session, we set up a gated mechanism to capture both long-term and short-term preference.
\begin{equation}
\mathbf{o}=w\log_\mathbf{o}^k(\mathbf{E}^{H,j}_{(n)})+(1-w)\log_\mathbf{o}^k(\mathbf{z}^{H}_{n}),
\end{equation}
where $\mathbf{E}^{H,j}_{(n)}$ denotes the embedding corresponding of the last item in the given session S.

Finally, after we get a unified preference representation $o$, we make a recommendation score for each element in the item set $V$.
\begin{equation}
\widehat{\mathbf{y}_i}=Softmax(\mathbf{o}^{T}\mathbf{v}_i),
\end{equation}
where $\widehat{\mathbf{y}_i}$ is the recommendation probability of our framework for item $v_i$. For the session-based recommendation task, we select the highest $K$ probabilities item from item set $V$  as final result according to $\widehat{\mathbf{y}_i}$.

\subsection{Contrastive Learning}
\label{contrastive_loss}
By projecting the item embedding into hyperbolic space, we empower the performance of our framework. In the recommendation scenario, there are many similar items, but users usually only choose their favorite items, so if we can let the model distinguish this subtle distinction, it may significantly improve the recommendation ranking performance of our framework. Inspired by the successful practice of contrastive learning, we  introduce contrastive learning in an innovative way into the framework in order to increase the modeling of user behavior. Compared with simple contrastive learning, the operation of our framework is calculating in hyperbolic space, which will be somewhat more complicated. Specifically, we want to make the best use of the distance between items in hyperbolic space through contrastive learning, then the recommendation model perceives more subtle distinction and improve the ranking performance.

We formulate our objective into two parts, the first part $L_e$ is cross-entropy loss function, which has been widely used in recommender system. The second part is the contrastive ranking loss $L_c$ with margin. The purpose of $L_c$ is to separate the positive and negative pairs up to a given margin. When the margin is reached, the pairs of items are considered to be properly segregated and with little loss. This enables the model keep focus on the pairs of items that are not near the margin and the margin separation is optimized in Lorentz hyperbolic space. \begin{equation}					
L_e=-\sum_{i=1}^{n}{({\mathbf{y}_i}\log{\left(\widehat{\mathbf{y}_i}\right)}+(1-{\mathbf{y}_i})\log{\left(1-\widehat{\mathbf{y}_i}\right)})} ,
\end{equation}
\begin{equation}
L_c=\sum_{i=1}^{n}{\max((d_L\left(\widehat{\mathbf{y}_i},\mathbf{y}_P\right)-d_L\left(\widehat{\mathbf{y}_i},\mathbf{y}_N\right)+\xi,0)},
\end{equation}
\begin{equation}
\label{loss}
L_{total}= \gamma\ast L_e+\beta\ast L_c
\end{equation}
where $\gamma$ and $\beta$ control the magnitude of the cross entropy loss and contrastive ranking loss respectively.

\section{Experiment}
\label{experiment}
In this section, detailed experiments will be conducted to assess the performance of the HCGR framework. We intend to answer following questions:

\begin{itemize}
\item \textbf{RQ1}: How does our proposed method perform comparing with the state-of-the-art methods?

\item \textbf{RQ2}: How the different components (i.e., Lorentz transportation, multi-hop graph aggregation and contrastive learning) affect the performance of HCGR?

\item \textbf{RQ3}: Can HCGR present reasonable explanation with regard to predicting user preference and get better recommendation results ?
\end{itemize}

In particular, we first describe the datasets and experimental configuration  (section \ref{exp_setup}). Then we compare the effectiveness of HCGR with several comparison methods (section \ref{overall_performance}). In section \ref{ablation_study}, we analyze in detail the generalization capability and the possibility of migration of the HCGR. Lastly, section \ref{case_study}, we set the case study and visualize the embedding in hyperbolic space (section \ref{emb_vis}).

\subsection{Experimental Setup}
\label{exp_setup}
\subsubsection{Datasets and Metrics}
We evaluate different recommenders based on four publicly available datasets, three of which are public benchmark datasets, i.e., $Yoochoose$, $Last.FM$, and $Ta$-$Feng$. Specifically, $Yoochoose$\footnote{\url{https://recsys.acm.org/recsys15/challenge/}} is the competition dataset of Recsys challenge 2015. It contains e-commerce website click within six months and related information. $Last.FM$ \footnote{\url{http://millionsongdataset.com/lastfm/}} dataset contains a set of users from $Last.FM$
online music service, which describes tagging and the music listening details of user. The $Ta$-$Feng$\footnote{\url{http://recsyswiki.com/wiki/Grocery shopping datasets}} dataset is a grocery dataset published by ACM RecSys, it covers goods ranging from food, office supplies to furniture. The fourth dataset is the financial service scenarios dataset $MYbank$, which is an industrial online recommendation platform in the Ant Group. $MYbank$ dataset describes users' interests and preferences in financial products such as debit, trust, accounting, which contains more than 5.6 million interactions from 691,701 users and brings more challenges compared with the three public datasets. The data statistical status after preprocessing is summarized in Table \ref{Tab:DataStatistic}, where Avg.I/user and Avg.I/item denote "average interaction per user" and "average interaction per item", respectively. To filter noisy data, we filter out items that appear less than 3 times, and then remove all user's behaviors less than 3 items on four datasets. After preprocessing, we split user behaviors into three parts, i.e., we randomly pick 80$\%$ as  training set, 10$\%$ as validation set for hyper-parameter tuning, and the remaining part for evaluating the performance of the model. Furthermore, to prevent overfitting, we set the patience argument to be 10 in the early stopping mechanism which denotes how many epochs we want to wait after the last time the validation metrics improved before breaking the training loop.

\begin{table}[]
\centering
\caption{statistics of datasets}
\begin{tabular}{ccccccccccc}
\toprule
\textbf{Dataset} & \textsf{Users} & \textsf{Items} & \textsf{Avg.I\textbf{/}user} & \textsf{Avg.I\textbf{/}item} & \textsf{Behaviors}\\

\midrule

\textbf{$Yoochoose$} & $136,456$ & $8,827$ & $4.45$ & $68.86$ & $0.68M$ \\
\textbf{$Last.FM$} & $1,876$ & $4,614$ & $41.31$ & $16.79$ & $0.077M$ \\
\textbf{$Ta$-$Feng$} & $29,131$ & $18,861$ & $27.67$ & $42.73$ & $0.81M$ \\
\textbf{$MYbank$} & $691,701$ & $3,188$ & $ 8.37$ & $ 1818.01$ & $5.79M$ \\
\bottomrule
\end{tabular}
\label{Tab:DataStatistic}
\end{table}

To fairly compare the generalization performance of each model, we evaluate for each user on his/her performance in the test set by adopting three recognized metrics: $HitRate$, $NDCG$ and $MRR$. Here, we choose $K=${$10,20$}  to show the different metrics for $HitRate@K$, $NDCG@K$ and $MRR@K$.

\begin{itemize}
\item \textbf{$HitRate@K:$} If one or more element of the label $y$ is shown in the prediction results $\hat{y}$, we call it a hit. The $HitRate$ is calculated as follow:
\begin{equation}
         HitRate@K=\frac{\sum_{s \in S} I\left(y^{s} \cap \hat{y^{s}} \neq \phi\right)}{|S|},\\
\end{equation}
where  $|\hat{y^{s}}|=K$ , $I(*)$ denotes the indicator function and $\phi$ is an empty set.
A larger value of $HitRate$ reflects the accuracy of the recommendation results.

\item  \textbf{$NDCG@K:$} Normalized Discounted Cumulative Gain $(NDCG)$ is a ranking based metric, which focuses on the order of retrieval results and is calculated in the following way:
\begin{equation}
        N D C G @ K=\frac{1}{N_{k}} \sum_{i=1}^{K} \frac{2^{I\left(\hat{y^{s}} \in y^{s} \right)}-1}{\log _{2}(i+1)},
\end{equation}
where $N_k$ is a constant to denote the maximum value of $NDCG@K$ given $|\hat{y^{s}}|$ and $I(*)$ denotes an indicator function.A large $NDCG$ value reflects a higher the ranking position of the expected item.

\item $MRR@K$ Mean Reciprocal Rank(MRR) when the r item is not in the higher $K$ position, the reciprocal is set to 0. It is formally given by:\\
\begin{equation}
    MRR@K=\frac{1}{|S|} \sum_{i=1}^{|S|}\frac{1}{\operatorname{rank}_{i}},
\end{equation}
where $rank_i$ denotes the position of the item in ${\hat{y^{s}}}$. $MRR$
is a normalize ranking take into account the order of recommendation list $y^{s}$. A large $MRR$ value reflects a higher ranking position of the expected item.
\end{itemize}

\subsubsection{Comparison Methods}

To demonstrate the performance of HCGR, we consider the following representative methods for performance comparisons:
\begin{itemize}
\item \textbf{FPMC} \cite{rendle2010factorizing} - a classical markov-based model, which considers the latest interaction.

\item \textbf{FOSSIL} \cite{he2016fusing}- a classical markov-based model, which captures personalized dynamics.

\item \textbf{GRU4Rec} \cite{hidasi2015session} - a representative RNN-based method for session-based recommendation, it stacks multiple GRU layers for session-parallel mini-batch training.

\item \textbf{NARM} \cite{li2017neural} - a hybrid encoder with attention mechanism to model  sequential behaviors in the current session.

\item \textbf{SASRec} \cite{kang2018self} - a self attention-based sequential recommender, which utilizes relatively few actions and considers long-range dependencies.

\item \textbf{STAMP} \cite{liu2018stamp} - a short term behavior priority attention-based method.

\item \textbf{SRGNN} \cite{wu2019session} - a graph-based recommender modeling session to learn item representations.

\item \textbf{GC-SAN} \cite{xu2019graph} - an improved version of SRGNN, which uses a GNN and multi-layered self-attention mechanism to compute the sequence level embeddings.

\item \textbf{FGNN} \cite{qiu2019rethinking} - a graph-based method, which uses a weight attention network to compute the graph level embeddings.

\item \textbf{LESSR} \cite{chen2020handling} - a session-based recommender with GNN, which innovatively utilizes auxiliary graph to generate item representation.

\item \textbf{HCGR} - our approach with novel attentive information aggregation, which utilize contrastive loss to optimize the model by considering the distance between positive and negative samples in hyperbolic space.

\end{itemize}

In this experiment, we set the maximum length of session to be $50$, and the embedding dimension to be $d$=128 for all datasets, the initial learning rate is uniformly set to 0.001 ,the linear schedule decay rate of 0.5 of every 3 epochs and $L_2$ penalty is $3e-3$. All parameters are initialized by Gaussian distribution with mean value of 0 and standard deviation of 0.1. The model cooperates with the Adam optimizer to complete the training.

\subsubsection{Data exploration}

\begin{figure}[h]
\centering
\subfigure[$Last.FM$]{
\begin{minipage}[t]{0.48\textwidth}
\includegraphics[width=7.5cm]{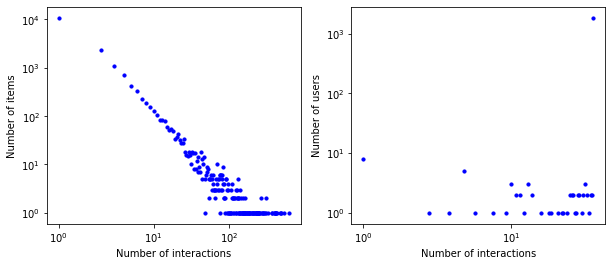}
\end{minipage}}
\subfigure[$Ta$-$Feng$]{
\begin{minipage}[t]{0.48\textwidth}
\includegraphics[width=7.5cm]{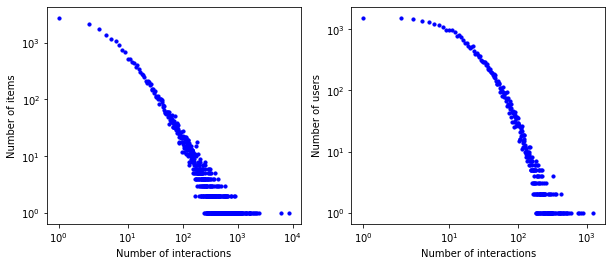}
\end{minipage}}
\subfigure[$Yoochoose$]{
\begin{minipage}[t]{0.48\textwidth}
\includegraphics[width=7.5cm]{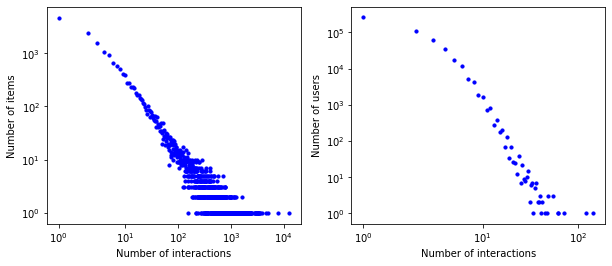}
\end{minipage}}
\subfigure[$MYbank$]{
\begin{minipage}[t]{0.48\textwidth}
\includegraphics[width=7.5cm]{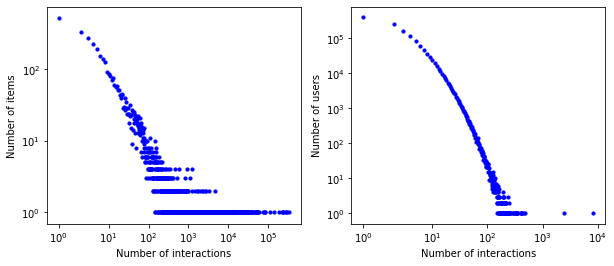}
\end{minipage}}
\caption{Distributions of user-item interaction, where (a-d) respectively show the results on four datasets. Corresponding to each dataset, the left one illustrates the distribution of the items clicked by each user and the right one plots the distribution of the users clicking each item.}
\label{lastfm34}
\end{figure}

As discussed in section \ref{hyperbolic_learning}, complex data with tree-like structure, i.e., the data obeys a power-law distribution, is effectively explained in the hyperbolic space. Therefore, we check the data distribution used in our experiment to verify the appropriateness of dataset selection. We present the distribution of number of interactions between users and items as illustrated in Figure \ref{lastfm34}. Power-law distribution is observed in three public datasets and the
$MYbank$ dataset. In the user-item interactions,  a majority of users interact with items very few times, meanwhile, most items are with few clicks. Such results demonstrate the tree-like structure of our dataset, which is supposed to have better representations in the hyperbolic space for session-based recommendation.

\subsection{Overall Performance (RQ1)}
\label{overall_performance}
\begin{table}[]
\small
\caption{Performance Comparisons}
\label{tab:experiment all}
\resizebox{\linewidth}{!}{
\begin{tabular}{@{}cccccccccccccc@{}}
\toprule
Dataset                     &Metric      & FPMC   & FOSSIL & GRU4Rec & NARM   & SASRec & STAMP  & SRGNN  & GC-SAN & FGNN   & LESSR  & \textbf{HCGR}   & \textbf{Improv.} \\ \midrule
\multirow{6}{*}{$Last.FM$}    & H@10 & 0.0623 & 0.0639 & 0.0759  & 0.0749 & 0.0808 & 0.0735 & 0.0902 & 0.0939 & 0.0946 & 0.106  & \textbf{0.1071} & \textbf{1.04\%}  \\
                           & M@10 & 0.03   & 0.0178 & 0.0288  & 0.0276 & 0.0342 & 0.0343 & 0.0411 & 0.0419 & 0.0414 & 0.0457 & \textbf{0.0523} & \textbf{14.44\%} \\
                           & N@10 & 0.0376 & 0.0286 & 0.0398  & 0.0386 & 0.0452 & 0.0411 & 0.0484 & 0.0542 & 0.0539 & 0.062  & \textbf{0.0651} & \textbf{5.00\%}  \\
                           & H@20 & 0.0923 & 0.0951 & 0.1131  & 0.1153 & 0.1192 & 0.1177 & 0.1099 & 0.1208 & 0.1241 & 0.1275 & \textbf{0.1388} & \textbf{8.86\%}  \\
                           & M@20 & 0.0321 & 0.02   & 0.0312  & 0.0303 & 0.037  & 0.0362 & 0.0431 & 0.0437 & 0.0434 & 0.0505 & \textbf{0.0541} & \textbf{7.13\%}  \\
                           & N@20 & 0.0452 & 0.0365 & 0.0491  & 0.0488 & 0.0549 & 0.0378 & 0.0354 & 0.0611 & 0.0614 & 0.0699 & \textbf{0.0749} & \textbf{7.15\%}  \\ \midrule
\multirow{6}{*}{$Yoochoose$} & H@10 & 0.4093 & 0.4014 & 0.4524  & 0.4615 & 0.4317 & 0.3967 & 0.4341 & 0.4768 & 0.4642 & 0.4735 & \textbf{0.4798} & \textbf{0.61\%}  \\
                           & M@10 & 0.1603 & 0.1471 & 0.2163  & 0.2207 & 0.1716 & 0.1915 & 0.2204 & 0.188  & 0.197  & 0.2241 & \textbf{0.2253} & \textbf{0.54\%}  \\
                           & N@10 & 0.219  & 0.2072 & 0.2719  & 0.2773 & 0.2328 & 0.2401 & 0.2709 & 0.2558 & 0.2598 & 0.2828 & \textbf{0.2898} & \textbf{2.48\%}  \\
                           & H@20 & 0.5013 & 0.4902 & 0.5544  & 0.5636 & 0.5391 & 0.4797 & 0.5279 & 0.5895 & 0.5687 & 0.5722 & \textbf{0.5938} & \textbf{0.73\%}  \\
                           & M@20 & 0.1668 & 0.1533 & 0.2235  & 0.2278 & 0.1791 & 0.1973 & 0.227  & 0.1959 & 0.2044 & 0.231  & \textbf{0.2325} & \textbf{0.65\%}  \\
                           & N@20 & 0.2424 & 0.2297 & 0.2978  & 0.3032 & 0.26   & 0.2611 & 0.2946 & 0.2844 & 0.2864 & 0.3078 & \textbf{0.3162} & \textbf{2.73\%}  \\ \midrule
\multirow{6}{*}{$Ta$-$Feng$}    & H@10 & 0.0853 & 0.0995 & 0.1091  & 0.1028 & 0.1091 & 0.0861 & 0.094  & 0.1099 & 0.1056 & 0.1115 & \textbf{0.1134} & \textbf{1.70\%}  \\
                           & M@10 & 0.04   & 0.0344 & 0.0456  & 0.0438 & 0.0447 & 0.0404 & 0.0435 & 0.0444 & 0.0396 & 0.0378 & \textbf{0.0487} & \textbf{28.84\%} \\
                           & N@10 & 0.0506 & 0.0497 & 0.0604  & 0.0576 & 0.0598 & 0.0511 & 0.0554 & 0.0587 & 0.0552 & 0.0533 & \textbf{0.0539} & \textbf{1.13\%}  \\
                           & H@20 & 0.1149 & 0.1358 & 0.1509  & 0.1401 & 0.1494 & 0.1181 & 0.1262 & 0.1403 & 0.1424 & 0.1477 & \textbf{0.1507} & \textbf{2.03\%}  \\
                           & M@20 & 0.042  & 0.0369 & 0.0485  & 0.0464 & 0.0475 & 0.0426 & 0.0458 & 0.0472 & 0.0422 & 0.0489 & \textbf{0.0512} & \textbf{4.70\%}  \\
                           & N@20 & 0.058  & 0.0589 & 0.0709  & 0.067  & 0.07   & 0.0592 & 0.0635 & 0.0699 & 0.0644 & 0.0673 & \textbf{0.0733} & \textbf{8.92\%}  \\ \midrule
\multirow{6}{*}{$MYbank$}    & H@10 & 0.5136 & 0.4521 & 0.5647  & 0.5459 & 0.5232 & 0.5542 & 0.5522 & 0.5505 & 0.5612 & 0.5562 & \textbf{0.5773} & \textbf{2.21\%}  \\
                           & M@10 & 0.2899 & 0.2623 & 0.3255  & 0.3185 & 0.3016 & 0.3167 & 0.3173 & 0.3164 & 0.3299 & 0.3104 & \textbf{0.3373} & \textbf{2.24\%}  \\
                           & N@10 & 0.3429 & 0.3073 & 0.3822  & 0.3724 & 0.354  & 0.373  & 0.373  & 0.3754 & 0.3835 & 0.3811 & \textbf{0.3901} & \textbf{1.72\%}  \\
                           & H@20 & 0.6185 & 0.5438 & 0.6647  & 0.6453 & 0.6261 & 0.6603 & 0.6544 & 0.6581 & 0.6684 & 0.6616 & \textbf{0.6713} & \textbf{0.43\%}  \\
                           & M@20 & 0.2972 & 0.2686 & 0.3324  & 0.3254 & 0.3087 & 0.3241 & 0.3245 & 0.3417 & 0.3445 & 0.3424 & \textbf{0.3578} & \textbf{3.86\%}  \\
                           & N@20 & 0.3694 & 0.3304 & 0.4075  & 0.3976 & 0.3799 & 0.3998 & 0.3989 & 0.4025 & 0.4103 & 0.4026 & \textbf{0.4135} & \textbf{0.78\%}  \\ \bottomrule
\multicolumn{13}{l}{* Realtive improvemens are calculated by comparing with the second best performance}
\end{tabular}
}
\end{table}

The experimental results of all comparison methods in session-based recommendation are presented in Table \ref{tab:experiment all}.
The best results of each column are highlighted in boldface. As can be observed, HCGR outperforms the best baselines with more than 4.5\% performance improvement on average on all datasets. From the results in Table \ref{tab:experiment all}, we can draw the following main findings:

\begin{itemize}
\item The RNN-based approaches which capture sequential dependency in a session(i.e., GRU4REC, NARM) remarkably outperform the traditional models that rely on Markov chains(i.e., FPMC, FOSSIL). This phenomenon proves that capturing sequential effects is a key factor for session-based recommendation as user's session-based behaviors are usually included in a short period and are likely to be temporally dependent.

\item  The attention-based models(i.e., NARM, SASRec, and STAMP) that involve attention mechanism get higher performance compared with that do not(i.e., GRU4REC) in all evaluation metrics. This is because NARM, STAMP, and SASRec can extract the shift of user interest within sessions and get the main purpose in the current session by incorporating an attention mechanism, which captures personal interest from the long-term memory or just models the user's current interest from the short-term behaviors. This phenomenon indicates that RNN-based approaches with the assumption that adjacent items in a session have a fixed sequential dependence may generate wrong dependencies, which further results in recommendation bias. This could be alleviated by involving the attention mechanism.

\item The GNN-based models (i.e., GC-SAN, FGNN) achieve better performance than RNNs-based models with or without attention mechanism due to the remarkable capacity of graph neural networks to capture complex interaction of user behaviors and describe the coherence of items in a session, which are ignored by RNNs-based models and such ignorance leads to overfitting in RNNs-based models.

\item Our proposed HCGR consistently outperforms all the comparison models on all datasets. Compared with FGNN and LESSR, our model involves an  advanced hyperbolic learning component to more effectively capture the coherence and hierarchy representations of the user behaviors within the Lorentz hyperbolic space, which ensures the correctness of the necessary representations' transformation. Furthermore, we use a novel graph message propagation mechanism with adaptive hyperbolic attention calculation to model user's preferences in session behavior sequences. In addition, we introduce contrastive learning to optimize the model by considering the distance between positive and negative samples in hyperbolic space, which can help learn better item representations.
\end{itemize}

\subsection{Ablation Study (RQ2)}
\label{ablation_study}
\subsubsection{Effect of Lorentz Transformation}

To demonstrate the effectiveness of the proposed hyperbolic learning framework for session-based recommendation, we conduct the ablation experiments by combining the Lorentz transformation with several baseline Euclidean SBR models, including FPMC, GRU4Rec, SASRec, and SRGNN. Besides, we also compare the performance of HCGR with ECGR (Euclidean Contrastive Graph Representation) by removing the Lorentz transformation from the hyperbolic contrastive graph representation learning framework shown in Figure \ref{Fig:NetworkArch}. The experimental results are shown
in Table \ref{tab:withlorentz}. The postfix $lorentz$ means the corresponding model is combined with hyperbolic learning to extract the hierarchy information contained within the SBR datasets. From Table \ref{tab:withlorentz}, we can draw the following conclusions:

\begin{itemize}
\item The performance of all models improves significantly on the four datasets when combining the Lorentz transformation with the Euclidean SBR models, which demonstrates that the hierarchy information from the power-law like session-based recommendation data is essential for predicting the user behavior, while such information is just ignored by the traditional SBR models built upon Euclidean space. Furthermore, the improvement of Markov-based method(i.e., FPMC) and attention-based method(i.e., SASRec) is more obvious than that of RNN-based(i.e., GRU4Rec) and GNN-based(i.e., SRGNN) method.

\item Our proposed hyperbolic contrastive graph representation learning method HCGR achieves the best results over all comparison models with or without Lorentz transformation, but the performance of ECGR drops evidently when replacing the Lorentz transformation with Euclidean transformation on all datasets. Besides, we find that the ECGR outperforms most baseline SBR models coupling with Lorentz transformation, which indicates the advantage of the proposed contrastive graph representation learning method.
\end{itemize}

\begin{table}[]

\small

\caption{Performance comparison of models with or without Lorentz on the $Last.FM$,  $Ta$-$Feng$, $Yoochoose$ and $MYbank$ datasets.}

\label{tab:withlorentz}

\resizebox{\linewidth}{!}{
\begin{tabular}{@{}cccccccccccc@{}}

\toprule

Dataset                    & Metric    & FPMC            & \begin{tabular}[c]{@{}c@{}}FPMC\_\\ lorentz\end{tabular} & GRU4Rec & \begin{tabular}[c]{@{}c@{}}GRU4Rec\_\\ lorentz\end{tabular} & SASRec          & \begin{tabular}[c]{@{}c@{}}SASRec\_\\ lorentz\end{tabular} & SRGNN           & \begin{tabular}[c]{@{}c@{}}SRGNN\_\\ lorentz\end{tabular} & ECGR           & \begin{tabular}[c]{@{}c@{}}HCGR\end{tabular} \\ \midrule

\multirow{6}{*}{$Last.FM$}   & H@10 & 0.0623          & \textbf{0.0684}                                         & 0.0759  & \textbf{0.0771}                                            & 0.0808          & \textbf{0.0868}                                           & 0.0902          & \textbf{0.0910}           & 0.1063                             & \textbf{0.1071}                               \\

                           & N@10    & 0.0376          & \textbf{0.0395}                                         & 0.0398  & \textbf{0.0400}                                            & 0.0452          & \textbf{0.0469}                                           & 0.0484          & \textbf{0.0495}        &  0.0642                            & \textbf{0.0651}                                         \\

                           & M@10     & 0.0300          & \textbf{0.0306}                                         & 0.0288  & \textbf{0.0288}                                            & 0.0342          & \textbf{0.0347}                                           & 0.0411          & \textbf{0.0417}            & 0.0483                             & \textbf{0.0523}                                     \\
                           & H@20 & \textbf{0.1149} & 0.0837                                                  & 0.1509  & \textbf{0.1526}                                            & 0.1494          & \textbf{0.1518}                                           & 0.1262          & \textbf{0.1349}             & 0.1294                             & \textbf{0.1388}                                   \\

                           & N@20    & \textbf{0.0452} & 0.0449                                                  & 0.0491  & \textbf{0.0509}                                            & 0.0549          & \textbf{0.0557}                                           & 0.0354          & \textbf{0.0588}              & 0.0718                             & \textbf{0.0749}                                   \\

                           & M@20     & \textbf{0.0321} & 0.0320                                                  & 0.0312  & \textbf{0.0317}                                            & 0.0370          & \textbf{0.0371}                                           & \textbf{0.0431} & 0.0425                       & 0.0516                             & \textbf{0.0541}                                   \\ \midrule

\multirow{6}{*}{$Ta$-$Feng$}   & H@10 & 0.0853          & \textbf{0.0863}                                         & 0.1091  & \textbf{0.1101}                                            & 0.1091          & \textbf{0.1110}                                           & 0.0940          & \textbf{0.0966}                        & 0.1068                             & \textbf{0.1134}                         \\

                           & N@10    & 0.0506          & \textbf{0.0555}                                         & 0.0604  & \textbf{0.0617}                                            & 0.0598          & \textbf{0.0603}                                           & 0.0554          & \textbf{0.0563}                & \textbf{0.0579}                             & 0.0539                                 \\

                           & M@10     & \textbf{0.0400} & 0.0392                                                  & 0.0456  & \textbf{0.0470}                                            & 0.0447          & \textbf{0.0448}                                           & 0.0435          & \textbf{0.0440}                 & 0.043                             & \textbf{0.0487}                                \\
                           & H@20 & 0.5013          & \textbf{0.5110}                                         & 0.5544  & \textbf{0.5577}                                            & 0.5391          & \textbf{0.5618}                                           & 0.5279          & \textbf{0.5442}                 & 0.1452                             & \textbf{0.1507}                    \\

                           & N@20    & \textbf{0.0580} & 0.0498                                                  & 0.0709  & \textbf{0.0724}                                            & 0.0700          & \textbf{0.0706}                                           & 0.0635          & \textbf{0.0659}                 & 0.0676                             & \textbf{0.0733}                                \\

                           & M@20     & \textbf{0.0420} & 0.0403                                                  & 0.0485  & \textbf{0.0499}                                            & 0.0475          & \textbf{0.0477}                                           & 0.0458          & \textbf{0.0466}                  & 0.0456                             & \textbf{0.0512}                               \\ \midrule

\multirow{6}{*}{$Yoochoose$} & H@10 & 0.4093          & \textbf{0.4149}                                         & 0.4524  & \textbf{0.4566}                                            & 0.4317          & \textbf{0.4463}                                           & 0.4341          & \textbf{0.4469}                      & 0.4651                             & \textbf{0.4798}                           \\

                           & N@10    & 0.2190          & \textbf{0.2243}                                         & 0.2719  & \textbf{0.2743}                                            & 0.2328          & \textbf{0.2359}                                           & \textbf{0.2709} & 0.2675                           & 0.2572                             & \textbf{0.2898}                               \\

                           & M@10     & 0.1603          & \textbf{0.1657}                                         & 0.2163  & \textbf{0.2181}                                            & \textbf{0.1716} & 0.1715                                                    & 0.2204          & \textbf{0.2221}                   & 0.1933                             & \textbf{0.2252}                              \\

                           & H@20 & 0.6185          & \textbf{0.6410}                                         & 0.6647  & \textbf{0.6685}                                            & \textbf{0.6261} & 0.6236                                                    & 0.6544          & \textbf{0.6649}                     & 0.5681                            & \textbf{0.5938}                            \\

                           & N@20    & 0.2424          & \textbf{0.2488}                                         & 0.2978  & \textbf{0.2999}                                            & 0.2600          & \textbf{0.2652}                                           & 0.2946          & \textbf{0.2952}                     & 0.2834                             & \textbf{0.3162}                           \\

                           & M@20     & 0.1668          & \textbf{0.1724}                                         & 0.2235  & \textbf{0.2252}                                            & 0.1791          & \textbf{0.1796}                                           & \textbf{0.2270} & 0.2253                             & 0.2005                             & \textbf{0.2325}                             \\ \midrule

\multirow{6}{*}{$MYbank$}    & H@10 & 0.5136       & \textbf{0.5372}                                         & 0.5647  & \textbf{0.5657}                                            & 0.5232          & \textbf{0.5253}                                           & 0.5522          & \textbf{0.5665}                           & 0.5607                             & \textbf{0.5772}                      \\

                           & N@10    & 0.3429          & \textbf{0.3552}                                         & 0.3822  & \textbf{0.3844}                                            & 0.3540          & \textbf{0.3617}                                           & 0.3730          & \textbf{0.3760}                    & 0.3823                             & \textbf{0.3901}                             \\

                           & M@10     & 0.2899          & \textbf{0.2987}                                         & 0.3255  & \textbf{0.3280}                                            & 0.3016          & \textbf{0.3112}                                           & 0.3173          & \textbf{0.3233}                    & 0.3114                              & \textbf{0.3373}                             \\

                           & H@20 & 0.6185          & \textbf{0.6410}                                         & 0.6647  & \textbf{0.6685}                                            & \textbf{0.6261} & 0.6236                                                    & 0.6544          & \textbf{0.6649}                     & 0.6619                             & \textbf{0.6713}                            \\

                           & N@20    & 0.3694          & \textbf{0.3814}                                         & 0.4075  & \textbf{0.4104}                                            & \textbf{0.3799} & 0.3766                                                    & 0.3989          & \textbf{0.4010}                      & 0.4079                             & \textbf{0.4135}                           \\

                           & M@20     & 0.2972          & \textbf{0.3059}                                         & 0.3324  & \textbf{0.3352}                                            & 0.3087          & \textbf{0.3181}                                           & 0.3245          & \textbf{0.3302}        & 0.3474                             & \textbf{0.3578}                \\
                \bottomrule
\end{tabular}
}
\end{table}

\begin{figure}[htbp]
\centering
\label{att_com}
\includegraphics[width=1\textwidth]{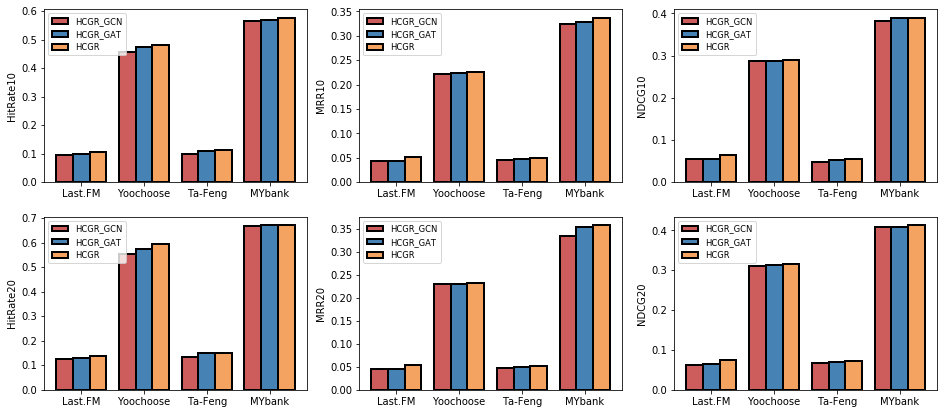}
\caption{The performance with different attention mechanisms on four datasets}
\label{att_com}
\end{figure}


\begin{table}[]
\caption{The performance with different optimization functions on $Last.FM$, $Yoochoose$ datasets.}
\label{loss_com_table}
\resizebox{\linewidth}{!}{
\begin{tabular}{@{}ccccccccccccc@{}}
\toprule
\multirow{2}{*}{Method} & \multicolumn{6}{c}{$Last.FM$} & \multicolumn{6}{c}{$Yoochoose$} \\
 & H@10 & M@10 & N@10 & H@20 & M@20 & N@20 & H@10 & M@10 & N@10 & H@20 & M@20 & N@20 \\ \cmidrule(r){1-13}
HCGR\_CE & 0.11 & 0.0508 & 0.0645 & 0.1358 & 0.052 & 0.0725 & 0.4751 & 0.2182 & 0.2861 & 0.5827 & 0.2261 & 0.3144 \\
HCGR & 0.1071 & 0.0523 & 0.0651 & 0.1388 & 0.0541 & 0.0749 & 0.4798 & 0.2252 & 0.2898 & 0.5938 & 0.2325 & 0.3162 \\
\textbf{Improv} & \textbf{-2.63\%} & \textbf{2.92\%} & \textbf{0.93\%} & \textbf{2.20\%} & \textbf{4.03\%} & \textbf{3.32\%} &\textbf{0.99\%}  & \textbf{3.21\%} & \textbf{1.29\%} & \textbf{1.90\%} &\textbf{2.83\%}  & \textbf{0.57\%} \\ \bottomrule
\end{tabular}
}
\label{loss_com_table}
\end{table}

\begin{table}[]
\caption{The performance with different optimization functions on $Ta$-$Feng$, $MYbank$ datasets.}
\resizebox{\linewidth}{!}{
\label{loss_com_table2}
\begin{tabular}{@{}ccccccccccccc@{}}
\toprule
\multirow{2}{*}{Method} & \multicolumn{6}{c}{$Ta$-$Feng$} & \multicolumn{6}{c}{$MYbank$} \\
 & H@10 & M@10 & N@10 & H@20 & M@20 & N@20 & H@10 & M@10 & N@10 & H@20 & M@20 & N@20 \\ \cmidrule(r){1-13}
HCGR\_CE & 0.1154 & 0.0453 & 0.0523 & 0.1532 & 0.048 & 0.0716 & 0.5743 & 0.3321 & 0.3876 & 0.665 & 0.3528 & 0.4123 \\
HCGR & 0.1134 & 0.0487 & 0.0539 & 0.1507 & 0.0512 & 0.0733 & 0.5722 & 0.3373 & 0.3901 & 0.6713 & 0.3578 & 0.4135 \\
\textbf{Improv} & \textbf{-1.70\%} & \textbf{7.5\%} & \textbf{3.06\%} & \textbf{-0.016\%} & \textbf{6.67\%} & \textbf{2.37\%} &\textbf{0.50\%}  & \textbf{1.56\%} & \textbf{0.64\%} & \textbf{0.95\%} &\textbf{1.47\%}  & \textbf{0.29\%} \\ \bottomrule
\end{tabular}
}
\label{loss_com_table2}
\end{table}

\subsubsection{Effect of Graph Aggregation Method}
To further investigate the advantage of the proposed adaptive hyperbolic graph aggregation method that utilizes multi-hop adjacent information, we conduct an ablation study by comparing different graph aggregation information approach within the framework of hyperbolic contrastive representation learning on four datasets. HCGR\_GCN refers to a model that the traditional spectrum-based graph convolution method is used to transport messages among adjacent neighbors, while HCGR\_GAT refers to a model that the graph attention-based method is used to aggregate adjacent information. The experimental results are shown in Figure \ref{att_com}. we can draw the following conclusions:

\begin{itemize}
\item The inductive attention-based graph convolution model(i.e., HCGR\_GAT) remarkably outperforms the transductive spectrum-based graph model(i.e., HCGR\_GCN) on all datasets, which indicates that treating neighbours deferentially and flexibly is essential to filter noisy information during message aggregation.

\item No surprisingly, HCGR with our proposed multi-hop adjacent information aggregation method achieves the best performance. Comparing to GAT, the main improvement of our proposed aggregation method relying on multi-hop aggregated message during graph node representation optimization is fully used, which indicates that low-order and high-order mutual graph information are both critical for final prediction. Such low-order mutual information is just ignored by GAT-like models.
\end{itemize}

\subsubsection{Effects of Contrastive Ranking Loss}
Diversity has become an important evaluation indices in recommendation scenario. In order to investigate the effectiveness of contrastive learning on the performance of our proposed hyperbolic graph representation learning framework, we conduct an experimental analysis by removing the contrastive ranking loss. To specific, HCGR\_CE means that contrastive ranking loss is removed from Eq (\ref{loss}) while keeping other settings same as HCGR. The experimental resutls are shown in Table \ref{loss_com_table} and Table \ref{loss_com_table2}, we can draw the following observations:

\begin{itemize}

\item In all datasets, the contrastive ranking loss optimization model HCGR outperforms the cross-entropy loss optimization model(HCGR\_CE) as regard to ranking evaluation metrics($MRR$, $NDCG$), which indicates that the contrastive ranking loss can distinguish the subtle distinction between items within sessions and improve recommendation diversity.

\item  With regard to the accuracy of recommendation results, there is no obvious difference between the performance of the two models, which indicates that our contrastive ranking loss can improve the ranking performances without losing recommendation accuracy.

\end{itemize}

\subsubsection{Effects Of Embedding Size}
\begin{figure}[htbp]
\centering
\includegraphics[width=1\textwidth]{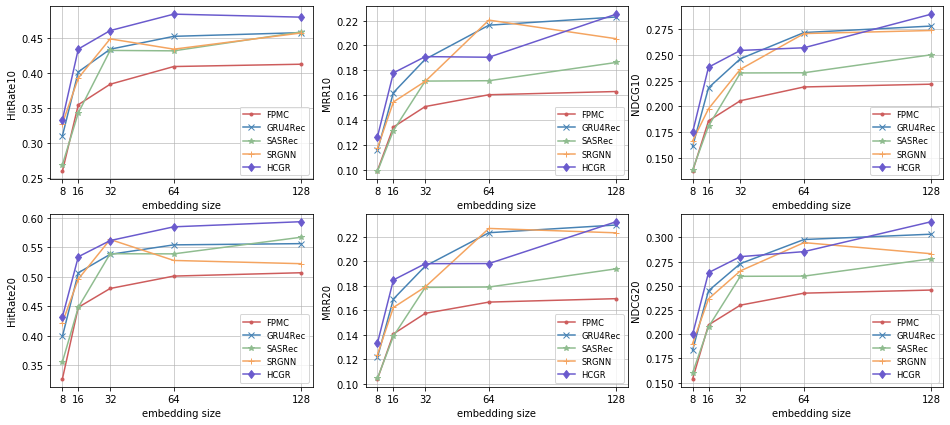}
\caption{The performance of various models with different embedding size on $Yoochoose$ datasets}
\label{yoo_emb}
\end{figure}

\begin{figure}[htbp]
\centering
\includegraphics[width=1\textwidth]{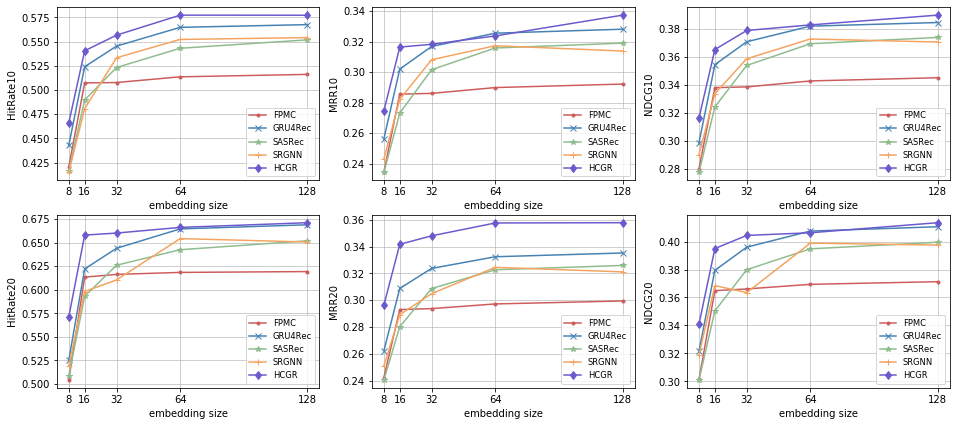}
\caption{The performance of various models with different embedding size on $MYbank$ datasets}
\label{alipay_emb}
\end{figure}

We explore the impact of embedding size $d$ on several evaluation indices as such size significantly affect the representation ability. We conduct the experiment on $Yoochoose$ and $MYbank$ datasets as their generalization and representation. The results are plotted in Figure \ref{yoo_emb}-\ref{alipay_emb}. We have the following observations:
\begin{itemize}
    \item HCGR outperforms all comparison SBR methods with most embedding sizes in all indices. Especially for small embedding size, such as 32, our model still can achieve better and robust results on these indices, which indicates introducing hyperbolic transformation can capture latent hierarchy property and boosts model performance.
    \item It is also observed that a proper embedding size is essential for graph node representation. When the embedding size is too small, it can't fully express node information and result in poor performance. On the opposite, a large embedding size may induce overfitting on the dataset.
\end{itemize}

\subsection{Case Study (RQ3)}
\label{case_study}
\subsubsection{Representation Analysis}
\begin{figure}[htb]
	\centering
	\includegraphics[width=1.0\linewidth]{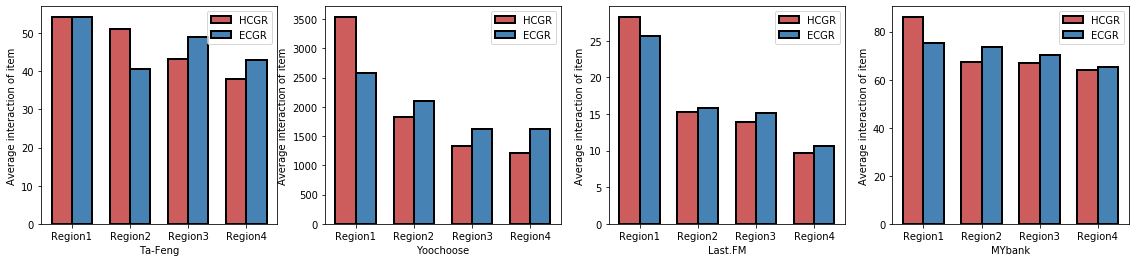}
    \caption{Hierarchical structure analysis of four datasets in tangent space and hyperbolic space}
	\label{case_study_1}
\end{figure}

We set up the case study to explore whether our model can learn the hierarchical structure of user behavior.
Whether the hierarchical structure in the data can be fully learned will affect the performance of the model, and this kind of hierarchical structure can be reflected by calculating the distance between the representation and the origin.
HCGR, ECGR are calculated in two different geometries, we use gyrovector space distance and tangent distance respectively to calculate the distance from the target point to the origin.
We set up three boundaries in Euclidean space and Lorentz space, and divide the representation into four regions according to their distance from the origin.
For example, the item of region 1 is the closest to the origin, whereas the item of region 4 is the farthest from the origin.
To intuitively reflect the different popularity of items in different regions, we count the interaction times of nodes in all regions of the four datasets.
We visualize the statistics as Figure ~\ref{case_study_1}.
From the results, it can be seen that the average number of interactions of items from region 1 to region 4 has decreased, which shows that both of our approaches ECGR and HCGR can model the hierarchical structure of session behavior.
In addition, in all datasets, the average number of interactions of items with HCGR in region 1 is higher than that with ECGR, while the average number of interactions of items with ECGR in region 3 and 4 is higher than that with HCGR.
Compared with ECGR, HCGR can better distinguish the items with different popularity and learn the hierarchical structure, which indicates that hyperbolic space is more suitable for embedding hierarchical data than Euclidean space for session-based recommendation tasks.

\begin{figure}[htb]
	\centering
	\includegraphics[width=1.0\linewidth]{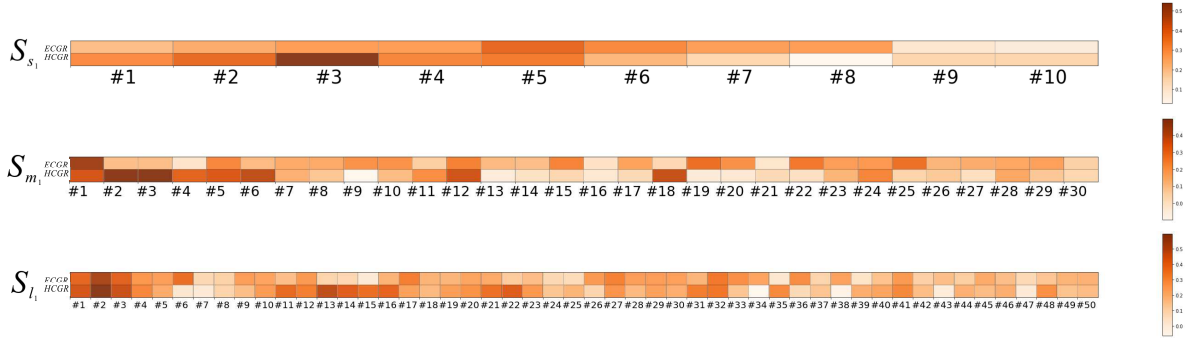}
    \caption{Visualization of average attention weight of behavior at different locations.}
	\label{case_study2}
\end{figure}

\begin{figure}[h]
    \centering
    \subfigure[$Last.FM$]{
    \begin{minipage}[b]{0.22\linewidth}
    \includegraphics[width=1\linewidth]{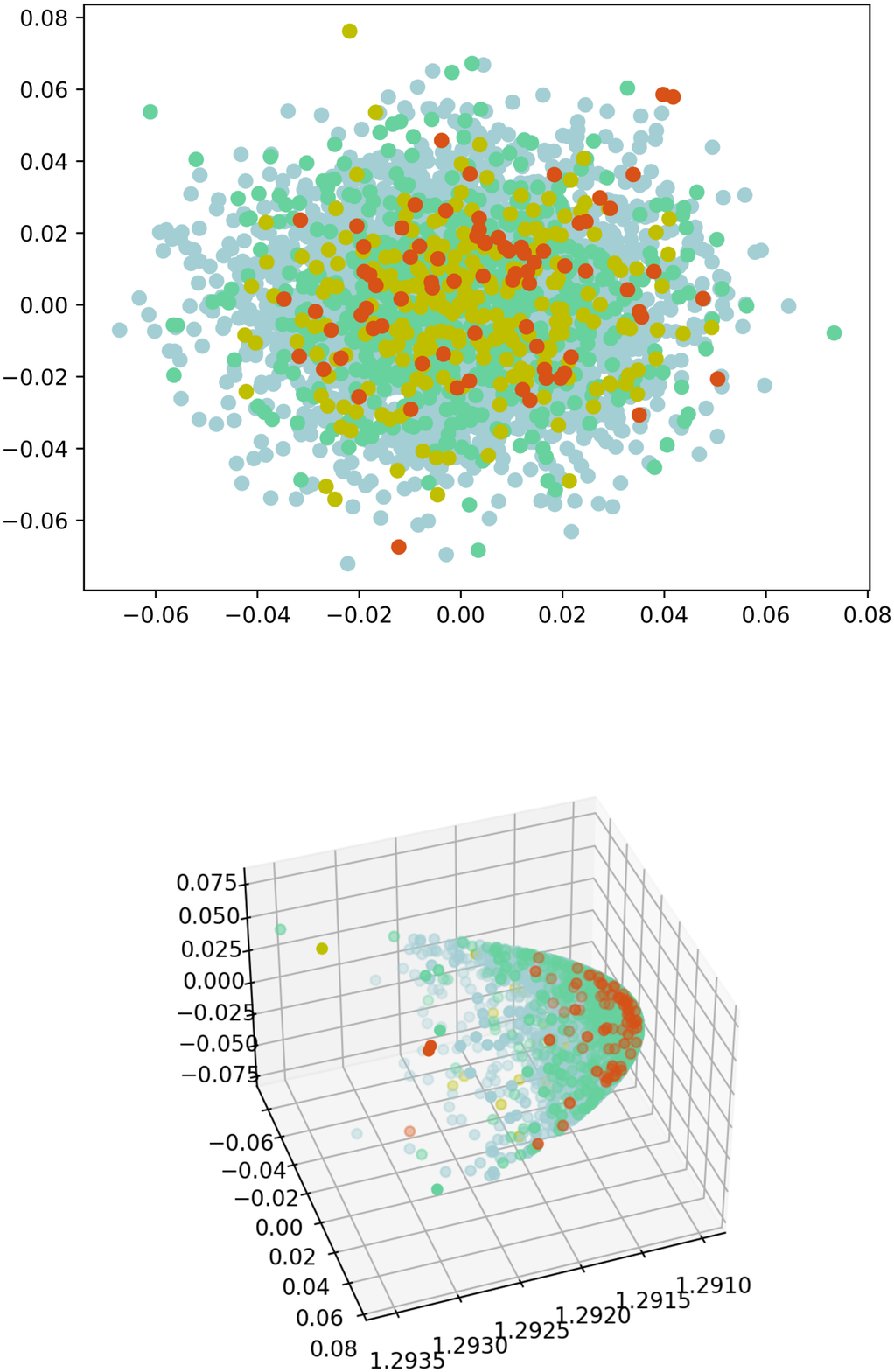}
    \end{minipage}}
    \subfigure[$Yoochoose$]{
    \begin{minipage}[b]{0.22\linewidth}
    \includegraphics[width=1\linewidth]{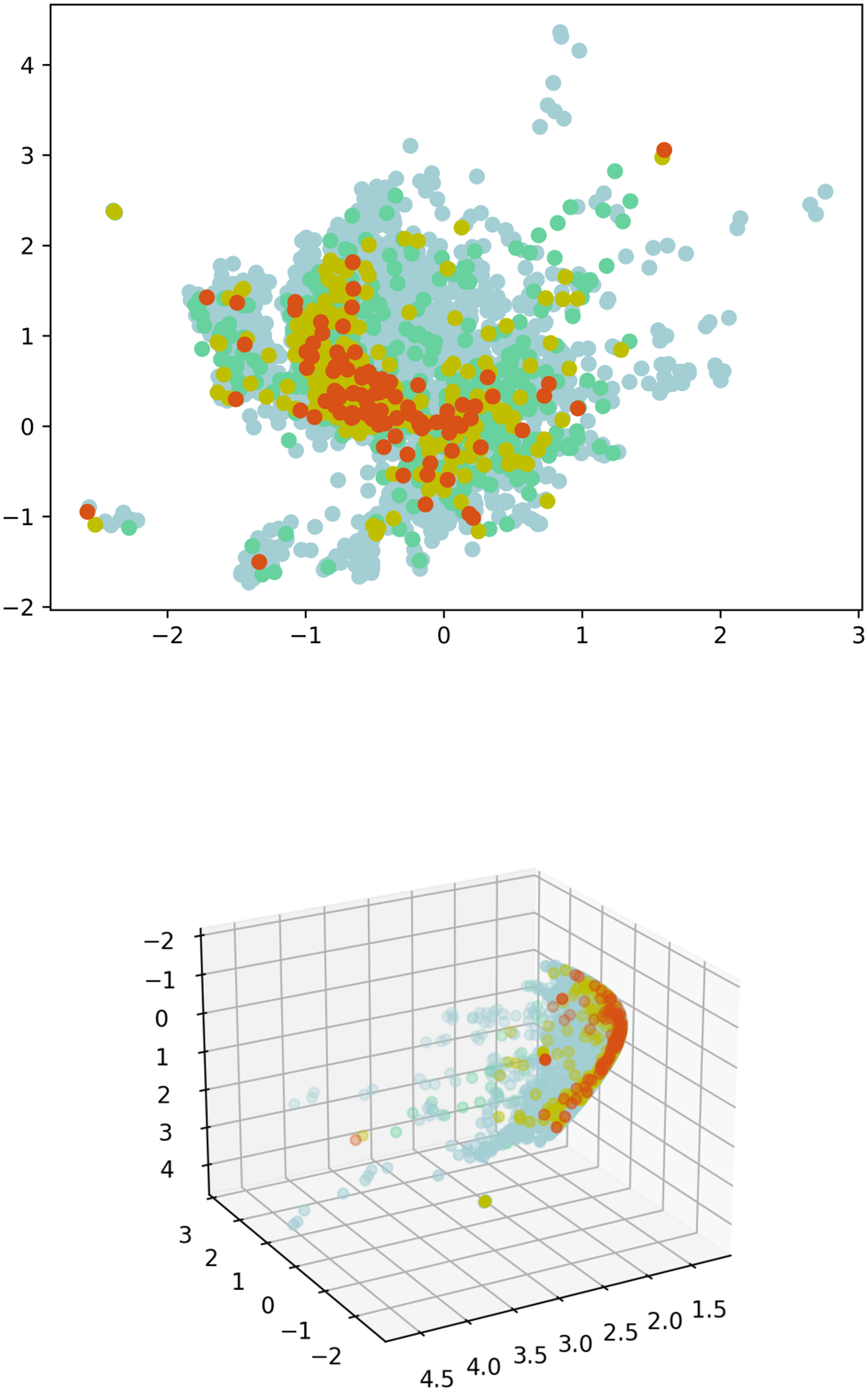}
    \end{minipage}}
    \subfigure[$Ta$-$Feng$]{
    \begin{minipage}[b]{0.22\linewidth}
    \includegraphics[width=1\linewidth]{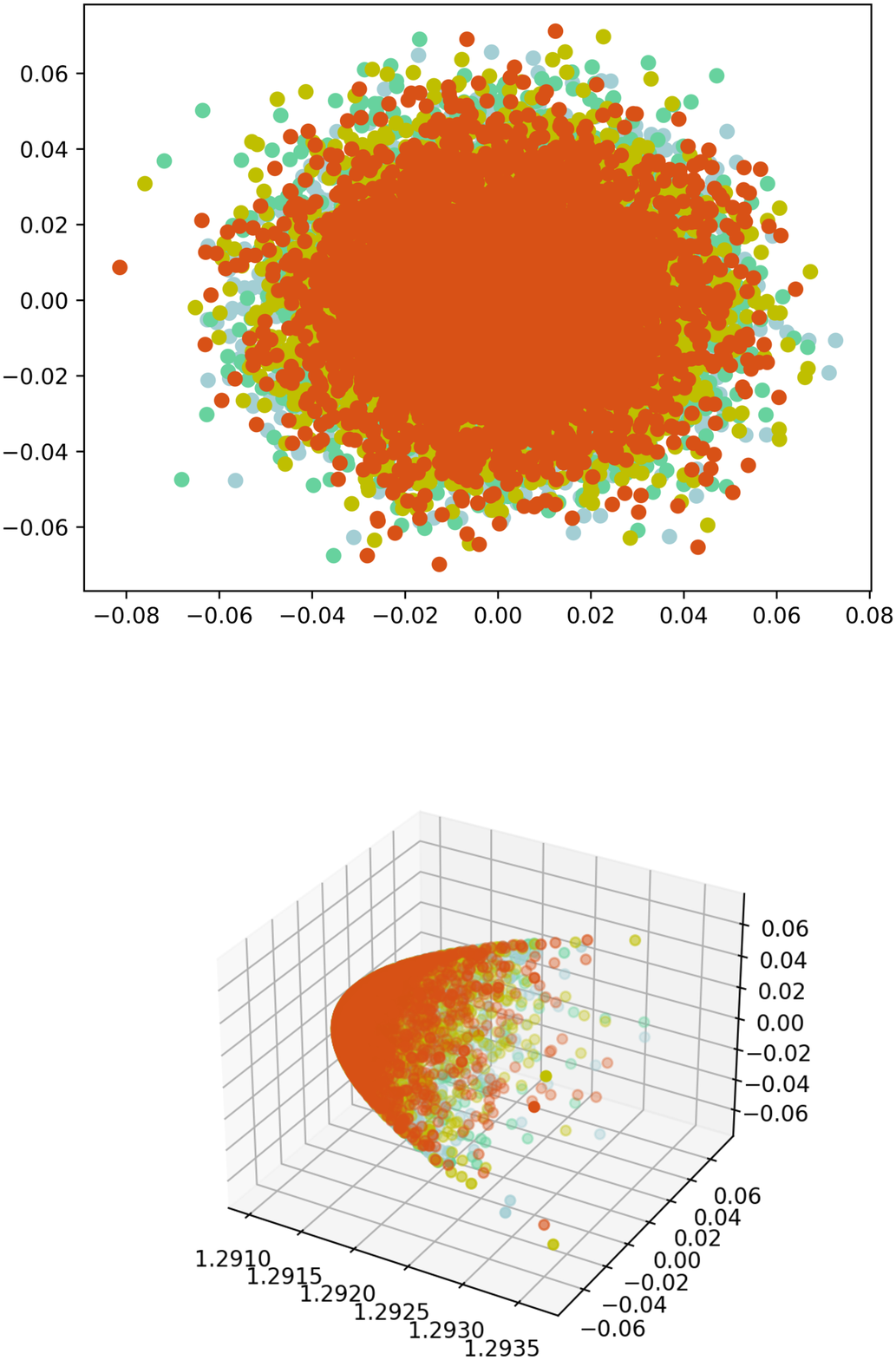}
    \end{minipage}}
    \subfigure[$MYbank$]{
    \begin{minipage}[b]{0.22\linewidth}
    \includegraphics[width=1\linewidth]{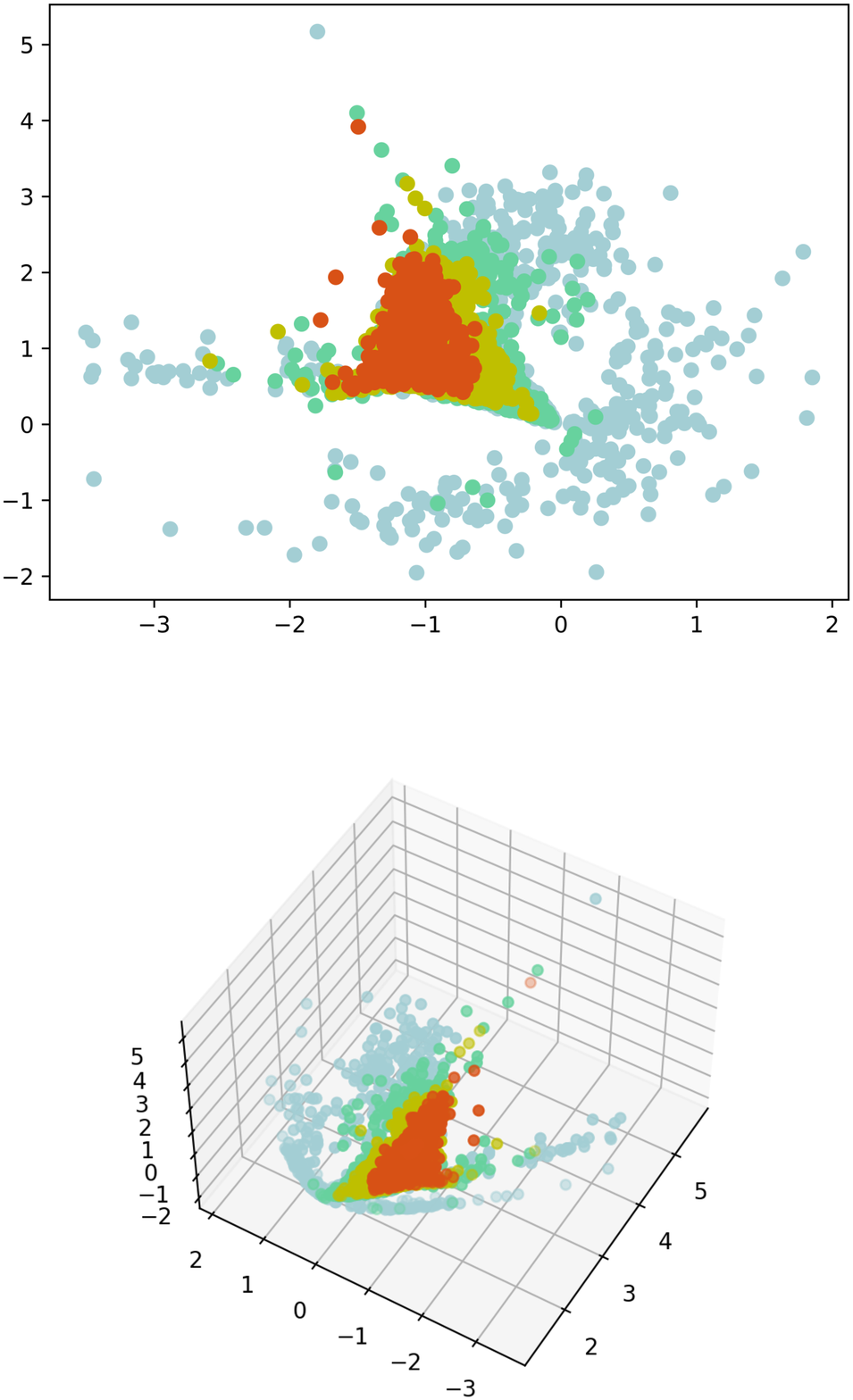}
    \end{minipage}}
    \caption{The embedding visualization, where (a-d) show the 2-d (in top row) and 3-d visualization (in bottom row) on four datasets resprectively.}
	\label{error-1}
\end{figure}

\subsubsection{Attention Analysis}
Taking advantage of the attention mechanism, we visualize the attention weight between user behaviors as shown in Figure ~\ref{case_study2}, which reflects the different influences within the same session on the two models (HCGR, ECGR). We randomly select three different sessions of length 10($S_{s_1}$), 30($S_{m_1}$) and 50($S_{l_1}$) respectively from $MYbank$ dataset(test set). For a same session in the heatmap, the above one is the attention weight between the related items and the next item user most likely to click modelled by ECGR, while the following one is the corresponding attention weight modelled by HCGR. From the heatmap, we discover that not all the behaviours in the same session equally contributing during the generation of the recommendation.
In addition, we also find that the attention weight of HCGR for session behavior is higher than that of ECGR in many key positions.
Specifically, HCGR will give higher scores with the increase of the scores given by ECGR, and HCGR can better distinguishes the item importance. It proves that hyperbolic space can better represent the hierarchical structure of data, thus making the attention mechanism capable of adaptively measuring the influence of session behavior.

\subsection{Embedding Analysis and Visualization}
\label{emb_vis}
We visualize the item embeddings in 2-dimension and 3-dimension on $Last.FM$, $Yoochoose$, $Ta$-$Feng$ and $MYbank$ datasets repectively according to Figure \ref{error-1}. (a) - (d). The popularity of the items is represented according to the different colors, decreasing with the color from red to green. Before training, we randomly initialize all the item embeddings. As shown in Figure \ref{error-1}, it is obvious that item embeddings present a hierarchical structure based on item popularity after training.
On the $Last.FM$ dataset, we can observe such a clear hierarchical representation, with the most popular items in the center and unpopular ones stay away from the center of projection space. Similar results can also be obtained on other datasets.

\section{Conclusion}
\label{conclusion}
The GNN-based model can not capture the hierarchical information effectively, which regularly appeared in recommendation scenarios. Enlightened by the powerful representation of non-Euclidean geometry which is proved to be able to reduce the distortion of embedding data onto power-law distribution, we proposed a hyperbolic contrastive graph recommender (HCGR), utilizing Lorentz hyperbolic space for item embeddings preserving their coherent and hierarchical properties. We design a novel hyperbolic graph message propagation mechanism due to the discrepancy between Euclidean and hyperbolic space during information passing. In addition, we introduce contrastive learning to enhance model performance by optimizing the distance between positive and negative samples of hyperbolic space, considering that distance in hyperbolic space can't be expressed well by traditional loss, such as CE, BPR loss. For future work, we will extend our method to the sequential recommendation which involves user profile and more item features. Besides, we will learn a parsimonious representation of symbolic data by embedding the dataset into spherical or product space and optimize the process of matrix multiplication in non-Euclidean geometry.

\bibliographystyle{unsrt}
\bibliography{sample-base}

\appendix

\end{document}